\documentclass [12pt,eqsecnum,amsfonts,aps]{revtex4}

\input epsf
\newcommand{\beq}{\begin{equation}}
\newcommand{\eeq}{\end{equation}}
\newcommand{\beqs}{\begin{eqnarray}}
\newcommand{\eeqs}{\end{eqnarray}}

\begin{document}

\baselineskip 6.0mm

\title{Exact Results on Potts Model Partition Functions in a
  Generalized External Field and Weighted-Set Graph Colorings} 

\bigskip

\author{Robert Shrock}
\email{robert.shrock@stonybrook.edu}

\author{Yan Xu}
\email{yan.xu@stonybrook.edu}

\affiliation{ C. N. Yang Institute for Theoretical Physics \\
State University of New York \\
Stony Brook, N. Y. 11794}


\begin{abstract}

We present exact results on the partition function of the $q$-state Potts model
on various families of graphs $G$ in a generalized external magnetic field that
favors or disfavors spin values in a subset $I_s = \{1,...,s\}$ of the total
set of possible spin values, $Z(G,q,s,v,w)$, where $v$ and $w$ are temperature-
and field-dependent Boltzmann variables. We remark on differences in
thermodynamic behavior between our model with a generalized external magnetic
field and the Potts model with a conventional magnetic field that favors or
disfavors a single spin value.  Exact results are also given for the
interesting special case of the zero-temperature Potts antiferromagnet,
corresponding to a set-weighted chromatic polynomial $Ph(G,q,s,w)$ that counts
the number of colorings of the vertices of $G$ subject to the condition that
colors of adjacent vertices are different, with a weighting $w$ that favors or
disfavors colors in the interval $I_s$.  We derive powerful new upper and lower
bounds on $Z(G,q,s,v,w)$ for the ferromagnetic case in terms of zero-field
Potts partition functions with certain transformed arguments. We also prove
general inequalities for $Z(G,q,s,v,w)$ on different families of tree graphs.
As part of our analysis, we elucidate how the field-dependent Potts partition
function and weighted-set chromatic polynomial distinguish, respectively,
between Tutte-equivalent and chromatically equivalent pairs of graphs.

\end{abstract}

\maketitle


\pagestyle{plain}
\pagenumbering{arabic}

\section{Introduction}

In this paper we continue our study of the $q$-state Potts model in a
generalized external magnetic field that favors or disfavors a certain subset
of spin values in the interval $I_s = \{1,...,s\}$, on various families of
graphs $G$ \cite{hl}-\cite{phs}.  We denote a graph $G=(V,E)$ by its vertex set
$V$ and its edge ( = bond) set $E$.  The numbers of vertices, edges, and
connected components of $G$ are denoted, respectively, by $n(G) \equiv n$,
$e(G)$, and $k(G)$. In thermal equilibrium at temperature $T$, the partition
function for the Potts model on the graph $G$ in this field is given by $Z =
\sum_{ \{ \sigma_i \} } e^{-\beta {\cal H}}$ with the Hamiltonian
\beq
{\cal H} = -J \sum_{\langle i j \rangle} \delta_{\sigma_i, \sigma_j}
- \sum_{p=1}^q H_p \sum_\ell \delta_{\sigma_\ell,p} \ ,
\label{ham}
\eeq
where $i, \ j, \ \ell$ label vertices of $G$, $\sigma_i$ are
classical spin variables on these vertices, taking values in the set
$I_q = \{1,...,q\}$, $\beta = (k_BT)^{-1}$, $\langle
i j \rangle$ denote pairs of adjacent vertices, $J$ is the spin-spin
interaction constant, and 
\beq
H_p = \cases{ H & for $1 \le p \le s$ \cr
              0 & for $s+1 \le p \le q$ } \ .
\label{hp}
\eeq
Thus, for positive (negative) $H$, the Hamiltonian favors (disfavors) spin
values in the interval $I_s$. This is a generalization of a conventional 
magnetic field, which would favor or disfavor one particular spin value. 
We denote $I_s^\perp$ as the orthogonal complement of $I_s$ in $I_q$, i.e.,
$I_s^\perp = \{s+1,...,q\}$, and we use the notation
\beq
K = \beta J \ , \quad h = \beta H \ , \quad y = e^K \ , \quad v = y-1 \ ,
\quad w=e^h \ .
\label{kdef}
\eeq
The physical ranges of $v$ are $v \ge 0$ for the Potts ferromagnet, and
$-1 \le v \le 0$ for the Potts antiferromagnet.

It is very useful to have a general graph-theoretic formula for $Z$ that does
not make any explicit reference to the spins $\sigma_i$ or the summation over
spin configurations, but instead expresses this function as a sum of terms
arising from the spanning subgraphs $G' \subseteq G$.  This formula was derived
and analyzed in Refs. \cite{ph,phs} and is
\beq
Z(G,q,s,v,w) = \sum_{G' \subseteq G} v^{e(G')} \
\prod_{i=1}^{k(G')} \, u_{n(G'_i)} 
\label{clusterws}
\eeq
where
\beq
u_m = q-s+sw^m = q+s(w^m-1)  \ . 
\label{um}
\eeq
This generalizes a spanning subgraph formula for $Z$ in the case $s=1$ due to
F. Y. Wu \cite{wu78,wurev}. In the special case $H=0$, Eq. (\ref{clusterws})
reduces to the cluster formula for the zero-field Potts model 
partition function \cite{fk}-\cite{baxterbook}, denoted $Z(G,q,v)$, namely 
\beq
Z(G,q,v) = \sum_{G' \subseteq G} v^{e(G')} \ q^{k(G')} \ . 
\label{fk}
\eeq
The original definition of the Potts model, (\ref{ham}), requires $q$ to be in
the set of positive integers ${\mathbb N}_+$ and $s$ to be a non-negative
integer. These restrictions are removed by Eq. (\ref{clusterws}).  Furthermore,
Eq. (\ref{clusterws}) shows that $Z$ is a polynomial in the variables $q$, $s$,
$v$, and $w$, hence our notation $Z(G,q,s,v,w)$.  If two graphs $G_1$ and $G_2$
are disjoint, then $Z(G_1 \cup G_2) = Z(G_1)Z(G_2)$ so, without loss of
generality, we will usually restrict to connected $G$ (although Eq. 
(\ref{clusterws}) leads to consideration of disconnected spanning subgraphs
$G'$).

An important special case is the zero-temperature antiferromagnet, $K =
-\infty$, i.e., $v=-1$, and we denote
\beq
Ph(G,q,s,w) \equiv Z(G,q,s,-1,w) \ . 
\label{phz}
\eeq
In this case the only contributions to $Z$ are those such that no two adjacent
spins have the same value.  Thus, $Ph(G,q,s,w)$ counts the number of proper
$q$-colorings of the vertices of $G$ with a vertex weighting that either
disfavors (for $0 \le w < 1$) or favors (for $w > 1$) colors in the interval
$I_s$. Here, a proper $q$-coloring is defined as an assignment of $q$ colors to
the vertices of a graph $G$ subject to the condition that no two adjacent
vertices have the same color.  We have denoted these coloring problems as DFSCP
and FSCP for \underline{d}is\underline{f}avored or \underline{f}avored
weighted-\underline{s}et graph vertex \underline{c}oloring \underline{p}roblems
\cite{phs}.  The associated set-weighted chromatic polynomial constitutes a
generalization of the conventional (unweighted) chromatic polynomial, which
counts the number of proper $q$-colorings of a graph $G$. Recent reviews of
chromatic polynomials include \cite{dktbook}-\cite{jsrev}.  

As is evident from the discussion above, the model defined by Eq. (\ref{ham})
with Eq. (\ref{hp}) is of interest both in the context of statistical mechanics
and in the context of mathematical graph theory.  It also has an application to
certain frequency allocation problems in electrical engineering \cite{phs}.

\section{Some Basic Properties of $Z(G,q,s,w,v)$ and $Ph(G,q,s,w)$}
\label{properties}

In this section we discuss some basic results about $Z(G,q,s,v,w)$ and
$Ph(G,q,s,w)$ that will be needed in our work.  Applying the factorization
\beq
w^m-1=(w-1)\sum_{j=0}^{m-1}w^j 
\label{wmfactorization}
\eeq
in Eq. (\ref{clusterws}) with $m=n(G_i')$, one sees that the variable $s$
enters in $Z(G,q,s,v,w)$, and $Ph(G,q,s,w)$ only in the combination
\beq
t=s(w-1) \ .
\label{tvar}
\eeq
Since $I_s \subseteq I_q$, whence $0 \le s \le q$, and since $w \ge 0$
for any real external field $H$, it follows that 
\beq
u_m = q+s(w^m-1) \ge 0 \ . 
\label{umpositive}
\eeq
Therefore, for the ferromagnetic case $v \ge 0$, each term in the sum over
spanning subgraphs in Eq. (\ref{clusterws}) is nonnegative.  For a given
spanning subgraph $G' \subseteq G$, consisting of a sum of $k(G')$ connected
components $G'_i$, where $i=1,...,k(G')$, the contribution to $Z(G,q,s,v,w)$ in
Eq. (\ref{clusterws}) is the number of spanning subgraphs $G'$ of a particular
topology, $N_{G'}$, times $v^{e(G')}\prod_{i=1}^{k(G')} u_{n(G'_i)}$, which has
the generic form 
\beq
N_{G'} \, v^{e(G')}\prod_{j=1}^{k(G')} \, u_{n(G'_j)} \ . 
\label{uform}
\eeq
Here 
\beq
\sum_{i=1}^{k(G')} n(G'_i) = n  \ . 
\label{sum_ngprime_i}
\eeq
Since some of the components $G'_i$ and $G'_j$ may have the same number of
vertices, $n(G'_i)=n(G'_j)$, the product in Eq. (\ref{uform}) can also be
written as $\prod_j (u_{r_j})^{p_j}$, where $r_j$ takes on certain values in
the set $\{1,...,n\}$ and the exponents $p_j$ are integers taking on certain
values in the set $\{1,...,k(G')\}$.  As a consequence of
Eq. (\ref{sum_ngprime_i}), these satisfy the relation
\beq
\sum_j p_j r_j = n \ . 
\label{rprel}
\eeq
Note that $u_m$ satisfies the identity 
\beq
u_m(q,s,w) = w^m \, u_m(q,q-s,w^{-1})
\label{umidentity}
\eeq
where we have written $u_m$ as a function of its three arguments $q, \ s, \ w$.
A given spanning subgraph $G'$ corresponds to a partition of the total set of
vertices depending on which edges are present and which are
absent.  The sum of the coefficients $N_{G'}$ of the various terms 
$N_{G'}\prod_j (u_{r_j})^{p_j}$ that multiply a given power $v^{e(G')}$ in Eq. 
(\ref{clusterws}) is ${e(G) \choose e(G')}$ since this is the number of ways of
choosing $e(G')$ edges out of a total of $e(G)$ edges.  These satisfy the
relation 
\beq
\sum_{e(G')=0}^{e(G)} {e(G) \choose e(G')} = 2^{e(G)} \ . 
\label{coeffsum}
\eeq
This reflects the fact that there are $2^{e(G)}$ spanning subgraphs of $G$, as
follows from the property that these are classified by choosing whether each
edge is present or absent, and there are $2^{e(G)}$ such choices.  In
mathematical graph theory, a loop is defined as an edge that connects a vertex
to itself and a cycle is a closed circuit along the edges of $G$.  In the
following we restrict to loopless graphs.  For any such $n$-vertex graph $G$,
the terms in $Z(G,q,s,v,w)$ proportional to $v^0$, $v^1$, and $v^{e(G)}$ can be
given in general, as
\beq
Z(G,q,s,v,w) = u_1^n + e(G)vu_2 u_1^{n-2} + ... + v^{e(G)}u_n \ . 
\label{zterms}
\eeq

The partition function $Z(G,q,s,v,w)$ satisfies the following identities 
\cite{hl}-\cite{phs} 
\beq
Z(G,q,s,v,1) = Z(G,q,0,v,w) = Z(G,q,v) \ , 
\label{zw1andzs0}
\eeq
(where, as above, $Z(G,q,v)$ is the zero-field Potts partition function), 
\beq
Z(G,q,s,v,w) = w^n \, Z(G,q,q-s,v,w^{-1}) \ , 
\label{zsym}
\eeq
(c.f. Eq. (\ref{umidentity})) and
\beq
Z(G,q,q,v,w) = w^n \, Z(G,q,v) \ .
\label{zsq}
\eeq
Setting $v=-1$ in these identities yields the corresponding relations for
$Ph(G,q,s,w)$; for example, Eq. (\ref{zsym}) yields
\beq
Ph(G,q,s,w) = w^n \, Ph(G,q,q-s,w^{-1}) \ .
\label{phsym}
\eeq

There are a number of equivalent ways of writing $Z(G,q,s,v,w)$ as sums of
powers of a given variable with coefficients depending on the rest of the
variables in the set $\{q,s,v,w\}$.  The basic spanning subgraph formula
(\ref{clusterws}) is a sum of powers of $v$.  A second convenient form in which
to express $Z(G,q,s,v,w)$ is as a sum of powers of $w$ with coefficients,
denoted as $\beta_{Z,G,j}(q,s,v)$, which are polynomials in $q$, $s$, and $v$:
\beq
Z(G,q,s,v,w) = \sum_{j=0}^n \beta_{Z,G,j}(q,s,v) \, w^j \ . 
\label{zsumw}
\eeq
The symmetry (\ref{zsym}) implies the following relation among the
coefficients: 
\beq
\beta_{Z,G,j}(q,s,v) = \beta_{Z,G,n-j}(q,q-s,v) \quad 
{\rm for} \ 0 \le j \le n \ . 
\label{zbetasym}
\eeq
For the special case $v=-1$, we write 
\beq
Ph(G,q,s,w) = \sum_{j=0}^n \beta_{G,j}(q,s) \, w^j \ , 
\label{phsumw}
\eeq
where 
\beq
\beta_{G,j}(q,s) \equiv \beta_{Z,G,j}(q,s,-1) \ . 
\label{betabeta}
\eeq
From (\ref{zbetasym}), we have 
\beq
\beta_{G,j}(q,s) = \beta_{G,n-j}(q,q-s) \quad {\rm for} \ 0 \le j \le n \ . 
\label{phbetasym}
\eeq
We have proved further that \cite{phs} 
\beq
\beta_{Z,G,n}(q,s,v)=Z(G,s,v)
\label{zbetajn}
\eeq
and 
\beq
\beta_{Z,G,0}(q,s,v)=Z(G,q-s,v) \ , 
\label{zbetaj0}
\eeq
so that for $v=-1$, $\beta_{G,0}(q,s)=P(G,q-s)$ and $\beta_{G,n}(q,s) =
P(G,s)$. Various general factorization results were also given in
Ref. \cite{phs} for these coefficients $\beta_{Z,G,j}(q,s,v)$ and
$\beta_{G,j}(q,s)$, including the following: 
\beq
{\rm For} \ 1 \le j \le n, \ \ \beta_{Z,G,j}(q,s,v) \ {\rm and} \
\beta_{G,j}(q,s) \ {\rm contain \ a \  factor \ of } \ s \ .
\label{zbetaj1n}
\eeq
\beq
{\rm For} \ 0 \le j \le n-1, \ \ \beta_{Z,G,j}(q,s,v) \ {\rm and} \ 
\beta_{G,j}(q,s) \ {\rm contain \ a \ factor} \ (q-s) \ .
\label{betalowerfactor}
\eeq
The minimum number of colors needed for a proper $q$-coloring of a graph $G$ is
the chromatic number, $\chi(G)$.  A further factorization property is that
\beq
\beta_{G,n}(q,s)\ {\rm contains \ the \ factor} \ \prod_{j=0}^{\chi(G)-1}(s-j) 
\ , 
\label{betanprod}
\eeq
and
\beq
\beta_{G,0}(q,s) \ {\rm contains \ the \ factor} \ 
\prod_{j=0}^{\chi(G)-1}(q-s-j) \ . 
\label{betaj0factors}
\eeq

A third useful type of expression for $Z(G,q,s,v,w)$ is 
\beq
Z(G,q,s,v,w) = \sum_{j=0}^n \alpha_{Z,G,n-j}(s,v,w) \, q^{n-j} \ .
\label{zsumq}
\eeq
With the notation 
\beq
\alpha_{G,n-j}(s,w) \equiv \alpha_{Z,G,n-j}(s,-1,w) \ ,
\label{alfalf}
\eeq
we then have 
\beq
Ph(G,q,s,w) = \sum_{j=0}^n \alpha_{G,n-j}(s,w) \, q^{n-j} \ .
\label{phsumq}
\eeq
This form is particularly convenient for comparisons with the conventional
unweighted chromatic polynomial $P(G,q) = Ph(G,q,0,w) = Ph(G,q,s,1)$. 

For a graph $G$, the number of linearly independent cycles, $c(G)$ (the
cyclotomic number), satisfies the relation
\beq
c(G) = e(G)+k(G)-n(G) \ . 
\label{crelation}
\eeq
A connected $n$-vertex graph with no cycles is a tree graph, $T_n$, while a
general graph with no cycles, which can be disconnected, is called a forest
graph. We denote a graph $G$ with no cycles as $G_{nc}$ and define
\beq
q' \equiv \frac{q}{s} \ , \quad v' \equiv \frac{v}{s} \ . 
\label{qvprime}
\eeq
In Ref. \cite{phs} we proved that for such a cycle-free graph $G_{nc}$,
\beq
Z(G_{nc},q,s,v,w) = s^n Z(G_{nc},q',1,v',w) \ .
\label{zscaled}
\eeq
This relation allows us to obtain $Z(G_{nc},q,s,v,w)$ from
$Z(G_{nc},q,1,v,w)$ for any cycle-free graph $G_{nc}$.  In particular,
all of the results for $Z(G,q,s,v,w)$ for various types of tree graphs
calculated in Ref. \cite{ph} for $s=1$ can be used to obtain the analogous
results for general $s$.

For a graph $G$, let us denote the graph obtained by deleting an edge $e \in E$
as $G-e$ and the graph obtained by deleting this edge and identifying the two
vertices that had been connected by it as $G/e$.  The Potts model partition
function satisfies the deletion-contraction relation (DCR)
\beq
Z(G,q,v) = Z(G-e,q,v)+vZ(G/e,q,v) \ , 
\label{zdelcon}
\eeq
and, setting $v=-1$, the chromatic polynomial thus satisfies the DCR
\beq
P(G,q)= P(G-e,q)-P(G/e,q) \ .   
\label{pdelcon}
\eeq
However, as we showed in Ref. \cite{phs}, in general, neither $Z(G,q,s,v,w)$
nor $Ph(G,q,s,w)$ satisfies the respective deletion-contraction relation, i.e.,
in general, $Z(G,q,s,w,v)$ is not equal to $Z(G-e,q,s,w,v)+vZ(G/e,q,s,w,v)$. 
The only cases where this deletion-contraction relation holds are for the
values $s=0$, $w=1$, and $w=0$ where $Z(G,q,s,v,w)$ reduces to a zero-field
Potts model partition function. In Section

\section{Upper and Lower Bounds on $Z(G,q,s,v,w)$ for $v \ge 0$}

In this section we derive powerful new two-sided upper and lower bounds for the
generalized field-dependent partition function of the ferromagnetic ($v \ge 0$)
Potts model, $Z(G,q,s,v,w)$ on an arbitrary graph $G$ in terms of the
zero-field Potts model partition functions $Z(G,u_1,v)$ and $Z(G,u_1/w,v)$,
where $u_1 = q+s(w-1)$ (c.f. Eq. (\ref{um})). These are especially useful
because the zero-field Potts model partition function is considerably easier to
calculate than $Z(G,q,s,v,w)$.  Throughout this section, it is understood that
$q \ge 0$, $0 \le s \le q$, and $v \ge 0$.  The former two conditions are
obvious for our present analysis, while the latter will often be indicated
explicitly.

We first derive a lower bound for $Z(G,q,s,v,w)$ for the range $w \ge 1$.  To
begin, we observe that, from its definition in Eq. (\ref{um}) and 
factorization property (\ref{wmfactorization}), $u_m$ satisfies
\beqs
u_m & = & q+s(w^m-1) = q+s(w-1)\sum_{j=0}^{m-1} w^j \cr\cr
   & \ge & q+s(w-1) = u_1 \quad {\rm for} \ \ w \ge 1 \ . 
\label{umgeu1}
\eeqs
Substituting this inequality into the expression for $Z(G,q,s,v,w)$ in 
Eq. (\ref{clusterws}) in terms of contributions from spanning subgraphs $G'
\subseteq G$, we have, for the same conditions
\beqs
Z(G,q,s,v,w) & = & \sum_{G' \subseteq G} v^{e(G')} \
\prod_{i=1}^{k(G')} \, u_{n(G'_i)} \cr\cr
& \ge & \sum_{G' \subseteq G} v^{e(G')} \, (u_1)^{k(G')} \quad {\rm for} \quad 
v \ge 0 \ \ {\rm and} \ \ w \ge 1 \ . 
\label{cluster1}
\eeqs
But the expression on the second line of Eq. (\ref{cluster1}) is just the
zero-field Potts model partition function given in Eq. (\ref{fk}) with its
argument $q$ replaced by $u_1$, namely $Z(G,u_1,v)$.  Hence, we have derived a
lower bound on $Z(G,q,s,v,w)$:
\beq
Z(G,q,s,v,w) \ge Z(G,u_1,v) \quad {\rm for} \ v \ge 0 \ \ 
{\rm and} \ \ w \ge 1 \ . 
\label{z_lowerbound_wge1}
\eeq

For the interval $0 \le w \le 1$, the inequality (\ref{umgeu1}) is
reversed: 
\beqs
u_m & \le u_1 \quad {\rm for} \ \ 0 \le w \le 1 \ , 
\label{umleu1}
\eeqs
and thus Eq. (\ref{cluster1}) is replaced by 
\beqs
Z(G,q,s,v,w) & \le & \sum_{G' \subseteq G} v^{e(G')} \, (u_1)^{k(G')}
 \quad {\rm for} \ v \ge 0 \ \ {\rm and} \ \ 0 \le w \le 1 \ . 
\label{cluster2}
\eeqs
Therefore, we obtain a second inequality, which is an upper bound:
\beq
Z(G,q,s,v,w) \le Z(G,u_1,v) \quad {\rm for} \ v \ge 0 \
{\rm and} \ 0 \le w \le 1 \ . 
\label{z_upperbound_wle1}
\eeq

To derive two-sided inequalities, we make use of the symmetry relation
(\ref{zsym}), which maps the interval $w \ge 1$ to the interval $0 \le w \le 1$
and vice versa.  Let us start with the case $w \ge 1$, for which we have
proved the lower bound (\ref{z_lowerbound_wge1}).  Now, from the symmetry
relation (\ref{zsym}) we know that $Z(G,q,s,v,w) = w^nZ(G,q,\hat s,v,\hat w)$
where $\hat s \equiv q-s$ and $\hat w \equiv w^{-1}$. Since $\hat w \in [0,1]$,
we can apply our upper bound (\ref{z_upperbound_wle1}) to 
$Z(G,q,\hat s,v,\hat w)$, getting the inequality 
\beq
Z(G,q,\hat s,v,\hat w) \le Z(G,\hat u_1,v) \ , 
\label{zineqaux1}
\eeq
where
\beq
\hat u_1 \equiv q+\hat s(\hat w - 1) = q+(q-s)(w^{-1}-1) = \frac{u_1}{w} \ . 
\label{hatu1}
\eeq
Combining (\ref{zineqaux1}) with (\ref{z_lowerbound_wge1}), we derive the
two-sided inequality
\beq
Z(G,u_1,v) \le Z(G,q,s,v,w) \le w^n \, Z(G, \frac{u_1}{w},v) \quad 
   {\rm for} \ v \ge 0 \ \ {\rm and} \ \ w \ge 1 \ .
\label{twosided_inequality_wge1}
\eeq
For the interval $0 \le w \le 1$, by the same type of reasoning, we extend our
upper bound (\ref{z_upperbound_wle1}) to the two-sided inequality 
\beq
w^n \, Z(G,\frac{u_1}{w},v) \le Z(G,q,s,v,w) \le Z(G,u_1,v) \quad 
   {\rm for} \ v \ge 0 \ \ {\rm and} \ \ 0 \le w \le 1 \ . 
\label{twosided_inequality_wle1}
\eeq
As two-sided inequalities, these are powerful restrictions on the
generalized field-dependent Potts model partition function in terms of
zero-field Potts model partition functions with $q$ replaced by $u_1$ and
$u_1/w$. 

We next prove some factorization properties of the upper and lower differences
in these two-sided inequalities. First, if $w=1$, then since
$Z(G,q,s,v,1)=Z(G,q,v)$ and $u_1=q$, it follows that the two-sided inequalities
(\ref{twosided_inequality_wle1}) and (\ref{twosided_inequality_wge1}) reduce to
equalities, i.e., both the upper and lower differences vanish.  Second, if
$v=0$, then the only contributions in the respective Eqs. (\ref{clusterws}) and
(\ref{fk}) are from the spanning subgraph with no edges (called the null graph,
$N_n$), so $Z(G,q,s,0,w)=(u_1)^n$, and $Z(G,q,0)=q^n$, whence $Z(G,u_1,0)=
(u_1)^n$ and $w^n \, Z(G,u_1/w,0) = (u_1)^n$.  Hence, again, in this
$v=0$ case, the inequalities (\ref{twosided_inequality_wle1}) and
(\ref{twosided_inequality_wge1}) reduce to equalities and the upper and lower
differences vanish. Third, if $s=0$, then $Z(G,q,0,v,w)=Z(G,q,v)$ and $u_1=q$,
so that $Z(G,u_1,v)=Z(G,q,v)$.  Hence, if $s=0$, then the lower difference in
(\ref{twosided_inequality_wge1}) and the upper difference in
(\ref{twosided_inequality_wle1}) vanish. Fourth, if $w=0$, then
$Z(G,q,s,v,0)=Z(G,q-s,v)$ and $u_1=q-s$, so $Z(G,u_1,v)=Z(G,q-s,v)$; therefore,
again, the lower difference in (\ref{twosided_inequality_wge1}) and the upper 
difference in (\ref{twosided_inequality_wle1}) vanish.  Together, these four 
results prove that the difference 
\beq
Z(G,q,s,v,w)-Z(G,u_1,v) \quad {\rm contains \ the \ factor} \quad w(w-1)sv \ . 
\label{lower_diff_factor_wge1}
\eeq
Fifth, if $s=q$, then $Z(G,q,q,v,w)=w^n \, Z(G,q,v)$ and $u_1=qw$, so $w^n \,
Z(G,u_1/w,v) = w^n \, Z(G,q,v)$.  Hence, if $s=q$, then the upper difference in
(\ref{twosided_inequality_wge1}) and the lower difference in
(\ref{twosided_inequality_wle1}) vanish.  Combining this with the first two
results above, we have shown that
\beq
w^n \, Z(G, \frac{u_1}{w},v) - Z(G,q,s,v,w) \quad {\rm contains \ the \ factor}
\quad (w-1)(q-s)v \ . 
\label{upper_diff_factor}
\eeq
It is also useful to characterize the difference between the zero-field Potts
model partition functions that constitute the upper and lower bounds in 
these two-sided inequalities (\ref{twosided_inequality_wge1}) and 
(\ref{twosided_inequality_wle1}).  For an arbitrary graph $G$, we have
\beq
w^nZ(G,\frac{u_1}{w},v) - Z(G,u_1,v) = \sum_{G' \subseteq G} v^{e(G')} \, 
(u_1)^{k(G')} \, \Big [ w^{n(G)-k(G')} - 1 \Big ] \ , 
\label{upper_lower_diff}
\eeq
where $G'$ is a spanning subgraph of $G$. Now the right-hand side of
Eq. (\ref{upper_lower_diff}) is nonzero only if $G$ has at least one edge, and,
in this case, the only nonvanishing contributions have $n(G) - k(G') \ge 1$. It
follows that
\beq
w^nZ(G,\frac{u_1}{w},v) - Z(G,u_1,v) \quad {\rm contains \ a \ factor} \quad 
v u_1 (w-1) \ . 
\label{upper_lower_diff_factor}
\eeq

It is worthwhile to give some illustrations of these two-sided inequalities
(\ref{twosided_inequality_wge1}) and (\ref{twosided_inequality_wle1}).  We
first do this for tree graphs.  For any $n$-vertex tree graph $T_n$, if $w \ge
1$, then the inequality (\ref{twosided_inequality_wge1}) reads
\beq
u_1(u_1 + v)^{n-1} \le Z(T_n,q,s,v,w) \le u_1(u_1+wv)^{n-1} 
\quad {\rm for} \quad v \ge 0 \ \ {\rm and} \ \ w \ge 1 \ . 
\label{twosided_inequality_tree_wge1}
\eeq
where we have used $Z(T_n,q,v)=q(q+v)^{n-1}$. If $w \in [0,1]$, then the
inequality (\ref{twosided_inequality_wle1}) reads
\beq
u_1(u_1 + wv)^{n-1} \le Z(T_n,q,s,v,w) \le u_1(u_1+v)^{n-1} 
\quad {\rm for} \quad v \ge 0 \ \ {\rm and} \ \ 0 \le w \le 1 \ . 
\label{twosided_inequality_tree_wle1}
\eeq
(This example also shows how the apparent singularity at $w=0$ arising from the
$u_1/w$ argument in $Z(G,u_1/w,v)$ on the left-hand side of the inequality
(\ref{twosided_inequality_wle1}) is removed by the $w^n$ factor, yielding a
nonsingular expression.)  One gains further insight by calculating the
differences between the polynomials that constitute the upper bound, the middle
term, $Z(T_n,q,s,w,v)$, and the lower bound for various tree graphs.  For the
path graph $L_2$ and $w \ge 1$, the differences that enter in the two-sided
inequality (\ref{twosided_inequality_tree_wge1}) are
\beq
u_1(u_1+wv) - Z(L_2,q,s,v,w) = (w-1)(q-s)v \ge 0
\label{upper_minus_zline2}
\eeq
and
\beq
Z(L_2,q,s,v,w) - u_1(u_1+v) = w(w-1)sv \ge 0 \ . 
\label{zline2_minus_lower}
\eeq
For $w \in [0,1]$ the differences that enter in
(\ref{twosided_inequality_tree_wle1}) are obvious reversals of these, viz., 
$u_1(u_1+v)-Z(L_2,q,s,v,w) = w(1-w)sv \ge 0$ and 
$Z(L_2,q,s,v,w) - u_1(u_1+wv) = (1-w)(q-s)v \ge 0$. 
For the path graph $L_3$ and $w \ge 1$, the differences in 
(\ref{twosided_inequality_tree_wge1}) are
\beq
u_1(u_1+wv)^2 - Z(L_3,q,s,v,w) = (w-1)(q-s)v\Big [ 2u_1 + v(w+1) \Big ] \ge 0
\label{upper_minus_zline3}
\eeq
and
\beq
Z(L_3,q,s,v,w) - u_1(u_1+v)^2 = w(w-1)sv\Big [ 2u_1 + v(w+1) \Big ] \ge 0 \ , 
\label{zline3_minus_lower}
\eeq
and similarly for $w \in [0,1]$.

Among $n$-vertex tree graphs, the star graph $S_n$ has a particularly simple
field-dependent Potts partition function, which was given in Ref. \cite{phs} 
and is derived by a direct evaluation of the general formula 
(\ref{clusterws}) (for any $v$): 
\beqs
Z(S_n,q,s,v,w) & = & 
\sum_{j=0}^{n-1} {n-1 \choose j} \, v^j \, u_{j+1} u_1^{n-1-j} 
\cr\cr
&=& (q-s)\Big [q+s(w-1)+v \Big ]^{n-1} + sw \Big [q+s(w-1)+wv \ 
\Big ]^{n-1} \ . \cr\cr 
& & 
\label{zstar}
\eeqs
Here $j$ is the number of edges in a given spanning subgraph $G'$, and the
numerical prefactor ${n-1 \choose j}$ in the first line of Eq. (\ref{zstar}) is
the number of ways of choosing $j$ edges out of the total number of edges,
$n-1$, in $S_n$.  For $v \ge 0$, substituting this result (\ref{zstar})
into the two-sided inequalities (\ref{twosided_inequality_tree_wge1}) and 
(\ref{twosided_inequality_tree_wle1}), we can derive general formulas for the
respective upper and lower differences.  If $w \ge 1$ we find, for the lower
difference in (\ref{twosided_inequality_tree_wge1}),
\beqs 
Z(S_n,q,s,v,w)-u_1(u_1+v)^{n-1} &=& sw\Big [(u_1+wv)^{n-1}-(u_1+v)^{n-1}\Big ] 
\cr\cr
& = & sw \sum_{j=0}^{n-1} {n-1 \choose j} \, (u_1)^{n-1-j} \, v^j \, (w^j-1) 
\cr\cr
& = & sw(w-1)v\sum_{j=1}^{n-1} {n-1 \choose j} \, (u_1)^{n-1-j} \, 
v^{j-1} \, \Big [ \sum_{\ell=0}^{j-1} w^\ell \Big ] 
\cr\cr
& & \ge 0 \ . 
\label{zstarlowerwge1}
\eeqs
In the same way, if $w \in [0,1]$, then the upper difference
$u_1(u_1+v)^{n-1}-Z(S_n,q,s,v,w)$ in (\ref{twosided_inequality_tree_wle1}) is
given by minus the right-hand side of Eq. (\ref{zstarlowerwge1}).  Similarly, 
if $w \ge 1$, then for the upper difference in 
(\ref{twosided_inequality_tree_wge1}) we calculate 
\beqs
u_1(u_1+wv)^{n-1}-Z(S_n,q,s,v,w) &=& (q-s)\Big [(u_1+wv)^{n-1}-
(u_1+v)^{n-1}\Big ] \cr\cr
& = & (q-s)(w-1)v\sum_{j=1}^{n-1} {n-1 \choose j} \, (u_1)^{n-1-j} \, 
v^{j-1} \, \Big [ \sum_{\ell=0}^{j-1} w^\ell \Big ] 
\cr\cr
& & \ge 0 \ . 
\label{zstarupperwge1}
\eeqs
Again, if $w \in [0,1]$, then the lower difference
$Z(S_n,q,s,v,w)-u_1(u_1+wv)^{n-1}$ in (\ref{twosided_inequality_tree_wle1}) is
given by minus the right-hand side of Eq. (\ref{zstarupperwge1}).

For the circuit graph $C_n$, if $w \ge 1$, the inequality 
(\ref{twosided_inequality_tree_wge1}) reads 
$Z(C_n,u_1,v) \le Z(C_n,q,s,v,w) \le w^n \, Z(C_n,u_1/w,v)$.  Using the fact
that $Z(C_n,q,v) = (q+v)^n+(q-1)v^n$, we can write this explicitly as 
\beq
(u_1+v)^n+(u_1-1)v^n \le Z(C_n,q,s,v,w) \le (u_1+wv)^n + (u_1-w)w^{n-1}v^n \ .
\label{twosided_inequality_circuit_wge1}
\eeq
For $C_2$ (which has a double edge), the differences that enter in this 
two-sided inequality are
\beq
w^2 \, Z(C_2,u_1/w,v) - Z(C_2,q,s,v,w) = (q-s)(w-1)v(v+2) \ge 0 
\label{upper_minus_zcirc2}
\eeq
and
\beq
Z(C_2,q,s,v,w) - Z(C_2,u_1,v) = w(w-1)sv(v+2) \ge 0 \ . 
\label{zcirc2_minus_lower}
\eeq
Similar illustrations of the general inequalities
(\ref{twosided_inequality_tree_wge1}) and (\ref{twosided_inequality_tree_wle1})
can be given for $L_n$ and $C_n$ with higher values of $n$ and for other 
families of graphs.

For $w \ge 1$, we can prove a lower bound on $u_m$ that is stronger than
(\ref{umgeu1}).  To do this, we use the basic inequality that for real positive
numbers $a_i$, the arithmetic mean is greater than or equal to the geometric
mean, i.e.,
\beq
\frac{1}{n} \, \sum_{j=1}^n a_j \ge \Big [ \prod_{j=1}^n a_j \Big ]^{1/n}
\label{arithmetic_geometric}
\eeq
(with equality only if $a_j = a_k \ \forall \ j, \ k$).  Applying this to the
sum $\sum_{j=0}^{m-1} w^j$ that appears in the factorization relation Eq. 
(\ref{wmfactorization}), we have, for all $w \ge 0$, 
\beq
\sum_{j=0}^{m-1} w^j \ge m \Big [ \prod_{j=0}^{m-1} w^j \Big ]^{1/m} \ . 
\label{wjinequality1}
\eeq
Now $\prod_{j=0}^{m-1} w^j = w^p$, where $p=\sum_{j=1}^{m-1} j$.  Using 
the summation formula $\sum_{j=1}^n j = n(n+1)/2$, we calculate that 
$p=(m-1)m/2$.  Hence, the inequality (\ref{wjinequality1}) for $w \ge 0$ 
can be written as 
\beq
\sum_{j=0}^{m-1} w^j \ge m w^{(m-1)/2} \ . 
\label{wjinequality2}
\eeq
Since $u_m = q+s(w^m-1) = q+s(w-1)\sum_{j=1}^{m-1} w^j$, we can use the lower
bound (\ref{wjinequality2}) to obtain a stronger lower bound on $u_m$ if $w \ge
1$ (but not if $w \in [0,1)$, since in that case the prefactor $(w-1)$ is
negative). Consequently, for $w \ge 1$, substituting (\ref{wjinequality2}) into
the expression for $u_m$, we derive the lower bound
\beq
u_m = q+s(w^m-1) \ge q+ms(w-1)w^{(m-1)/2} \quad {\rm for} \ \ w \ge 1 \ . 
\label{umlowerbound}
\eeq
Clearly, this is an improvement over the lower bound (\ref{umgeu1}).
Substituting this result into Eq. (\ref{clusterws}) with $u_m=u_{n(G'_i)}$, 
we thus obtain the following improved lower bound on $Z(G,q,s,v,w)$ for 
$w \ge 1$ and the ferromagnetic range $v \ge 0$ (where $G'$ is a spanning 
subgraph of $G$): 
\beqs
& & Z(G,q,s,v,w) \ge \sum_{G' \subseteq G} v^{e(G')} \
\prod_{i=1}^{k(G')} \, \Big [ q+n(G'_i)s(w-1)w^{(n(G'_i)-1)/2} \Big ] 
\quad {\rm for} \ \ w \ge 1 \ \ {\rm and} \ \ v \ge 0 \ . \cr\cr
& & 
\label{zineq1}
\eeqs
Note, however, that in contrast with our previous lower bound (\ref{cluster1}),
the right-hand side of this inequality cannot, in general, be written in terms
of a zero-field Potts model partition function since the terms in the product
depend explicitly on $n(G'_i)$.

\section{Some Thermodynamic Properties} 

The zero-field Potts model Hamiltonian ${\cal H}$ and partition function $Z$
are invariant under the global transformation in which $\sigma_i \to g \sigma_i
\ \ \forall \ \ i \in V$, with $g \in {\cal S}_q$, where ${\cal S}_q$ is the
symmetric (= permutation) group on $q$ objects.  In the presence of the
generalized external field defined in Eq. (\ref{hp}), this symmetry group of
${\cal H}$ and $Z$ is reduced to the tensor product
\beq
{\cal S}_q \to {\cal S}_s \otimes {\cal S}_{q-s} \ . 
\label{symgroup}
\eeq
This simplifies to the conventional situation in which the external field $H$
favors or disfavors only a single spin value if $s=1$ or $s=q-1$, in which case
the right-hand side of Eq. (\ref{symgroup}) is ${\cal S}_{q-1}$.  For $s$ in
the interval
\beq
2 \le s \le q-2 \ , 
\label{srange}
\eeq
the general model of Eqs. (\ref{ham}) and (\ref{hp}) exhibits properties that
are interestingly different from those of a $q$-state Potts model in a
conventional magnetic field.  For example, in the conventional case, at a given
temperature $T$, if $H >> |J|$, the interaction with the external field
dominates over the spin-spin interaction, and if $h = \beta H$ is sufficiently
large, the spins tend to be frozen to the single favored value.  In contrast,
here, at a given temperature $T$, provided that $s$ lies in the interval
(\ref{srange}), if $|H| >> |J|$, this effectively reduces the model to (i) an
$s$-state Potts model if $H > 0$, or (ii) a $(q-s)$-state Potts model if $H <
0$. In this limit, for given values of $q$ and $s$ and a given graph (say a
regular lattice), there are thus, in general, four types of possible models,
depending on both the sign of $H$ and the sign of $J$.  As an illustration of
this, let us consider the case $q=5$, $s=2$ on (the thermodynamic limit of) a
square lattice.  For $H=0$, the ferromagnetic version of the model has a
first-order phase transition, with spontaneous breaking of the ${\cal S}_5$
symmetry, at $K_c = \ln(1+\sqrt{5}) \simeq 1.17$, while the antiferromagnetic
version has no finite-temperature phase transition and is disordered even at
$T=0$ \cite{wurev,baxterbook}. For $H > 0$ and $H >> |J|$, the theory reduces
effectively to a two-state Potts model, i.e., an Ising model.  Because the
square lattice is bipartite, there is an elementary mapping that relates the
ferromagnetic and antiferromagnetic versions of the model, and, as is well
known, both have a second-order phase transition, with spontaneous symmetry
breaking of the ${\cal S}_2 \approx {\mathbb Z}_2$ symmetry, at
$|K_c|=\ln(1+\sqrt{2}) \simeq 0.881$ (where $K=\beta J$), with thermal and
magnetic critical exponents $y_t=1$, $y_h=15/8$, described by the rational
conformal field theory (RCFT) with central charge $c=1/2$.  For $H < 0$ and
$|H| >> |J|$, the theory effectively reduces to a three-state Potts model.  In
the ferromagnetic case, $J > 0$, this has a well-understood second-order phase
transition, with spontaneous symmetry breaking of the ${\cal S}_3$ symmetry, at
$K_c = \ln(1+\sqrt{3}) \simeq 1.01$, with thermal and critical exponents
$y_t=6/5$, $y_h=28/15$, described by a RCFT with central charge $c=4/5$
\cite{wurev,baxterbook,cft}.  In the antiferromagnetic case, $J < 0$, the model
has no finite-temperature phase transition but is critical at $T=0$ (without
frustration), with nonzero ground-state entropy per site $S/k_B = (3/2)\ln(4/3)
\simeq 0.432$ \cite{wurev,lieb}.

In particular, an interesting difference with respect to the $q$-state Potts
model with a conventional external magnetic field appears in the case in which
the spin-spin interaction is antiferromagnetic, i.e., $J < 0$.  In the
conventional case, there is competition between the two terms in the
Hamiltonian, and resultant frustration.  Here the situation is altered and
depends on the chromatic number $\chi(G)$ of the graph.  If $H > 0$, then there
is frustration if $s < \chi(G)$, and this becomes increasingly severe as the
temperature decreases, but if $s \ge \chi(G)$, then this frustration is absent,
because it is possible to satisfy the antiferromagnetic short-range ordering
preferred by the spin-spin interaction while also satisfying the assignments of
spin values preferred by the interaction of spins with the external field. (Of
course, the presence of this field does have an effect in restricting the
preferred range of values of the spins.)  Similarly, if $H < 0$, then there is
frustration if $(q-s) < \chi(G)$ but not if $(q-s) \ge \chi(G)$.  As an
example, we may consider the case $q=5$, $s=2$ on (the thermodynamic limit of)
a triangular lattice.  For $H > 0$ with $H >> |J|$, the model reduces to an
Ising model, and (i) if $J > 0$, this has a symmetry-breaking second-order
phase transition at $K_c=(1/2)\ln 3 \simeq 0.549$, in the same universality
class as on the square lattice, while (ii) if $J < 0$, there is frustration
and, as a consequence, the model has no finite-temperature phase transition,
but is critical at $T=0$, with nonzero ground-state entropy $S/k_B \simeq
0.323$ \cite{wannier}.  For $H < 0$ with $|H| >> |J|$, the model reduces to a
three-state Potts model, and (iii) if $J > 0$, this has a symmetry-breaking
second-order phase transition at $K_c=\ln[\cos(2\pi/9)+\sqrt{3} \,
\sin(2\pi/9)] \simeq 0.631$ \cite{kimjoseph}, in the same universality
class as on the square lattice; while (iv) if $J < 0$, it has a weakly
first-order symmetry-breaking phase transition at $K_c \simeq -1.59$
\cite{adler,hcl}, with a completely ordered ground state, reflecting the fact
that the chromatic number of the triangular lattice is $\chi(tri)=3$.  The more
general case where $|H|$ is not $>> |J|$ encompasses a rich variety of
thermodynamic behavior depending on the signs of $H$ and $J$, the ratio of
$|H/J|$, the values of $q$ and $s$, the dimensionality of the lattice, and, in
the antiferromagnetic case, the type of $d$-dimensional lattice.  Note that if
$J=0$, then (i) $S/k_B = \ln s$ for $H > 0$, and (ii) $S/k_B = \ln(q-s)$ for $H
< 0$.

Although a one-dimensional spin system (with short-ranged spin-spin
interactions, as is the case here) does not exhibit any finite-temperature
phase transition, it can still serve as a worthwhile illustration of some
thermodynamic properties.  A simple example of this type is provided by our
model on an infinite one-dimensional lattice, with either free or periodic
boundary conditions. We denote a reduced, dimensionless free energy per site as
$f = \lim_{n \to \infty} (1/n)\ln Z$. Then from our analysis above, we have,
for the limits as $n \to \infty$ of the line and circuit graphs, $\{L\}$ and
$\{C\}$,
\beq
f(\{L\},q,s,v,w) = f(\{C\},q,s,v,w) \equiv f_{1D}(q,s,v,w)
 = \ln(\lambda_{Z,1,0,+}) \ , 
\label{fline}
\eeq
where $\lambda_{Z,1,0,+}$ is given below in Eq. (\ref{lamcnplusminus}). From
this the various thermodynamic quantities such as the internal energy, specific
heat, entropy, etc. can be calculated.  Also, from this, one can obtain the
function $\Phi(\{G\},q,s,w) = \lim_{n \to \infty} Ph(G_n,q,s,w)^{1/n}$ for the
$n \to \infty$ limits of $G_n = L_n, \ C_n$. The function $\Phi(\{G\},q,s,w)$
generalizes the ground state degeneracy per site of the zero-temperature Potts
antiferromagnet, $W(\{G\},q) = \lim_{n \to \infty} P(G,q)^{1/n}$. We have
\beq
\Phi(\{L\},q,s,w) = \Phi(\{C\},q,s,w) \equiv \Phi_{1D}(q,s,w) = 
(\lambda_{Z,1,0,+})|_{_{v=-1}} \ . 
\label{philine}
\eeq
We note the following reductions of $\Phi_{1D}$, which follow from the
general identities given above:
\beq
\Phi_{1D}(q,0,w) = \Phi_{1D}(q,s,1) = q-1 \ ,
\label{phi1d_s0w1}
\eeq
\beq
\Phi_{1D}(q,s,0) = q-s-1 \ , 
\label{phi1d_w0}
\eeq
and
\beq
\Phi_{1D}(q,q,w) = w(q-1) \ . 
\label{phi1d_sq}
\eeq
With the ranges $0 \le s \le q$ and $w \ge 0$ understood,, $\partial
\Phi_{1D}/\partial q \ge 0$ for the nontrivial interval $q \ge 2$, reflecting
the greater freedom of color assignments with increasing $q$. Furthermore,
$\partial \Phi_{1D}/\partial w \ge 0$, as is clear from the original
Hamiltonian formulation in Eqs. (\ref{ham}) and (\ref{hp}).  The derivative
$\partial \Phi_{1D}/\partial s \ge 0$ if $w \ge 1$, and $\partial
\Phi_{1D}/\partial s \le 0$ if $0 \le w \le 1$, which follows from the fact
that the external field favors (disfavors) spin values in $I_s$ if $w > 1$ ($w
\in [0,1)$). Plots of $\Phi_{1D}$ as a function of $q$ and $w$ for fixed $s$
are similar to the $s=1$ results shown in Figs. 2-4 of Ref. \cite{ph}, except
that the minimal value of $q$ allowed is now $s$ instead of 1, and the line for
$w=0$ is now $\Phi(\{L\},q,s,0)=q-s-1$ rather than $q-2$.  We proceed to give
exact results for $Z(G,q,s,v,w)$ and $Ph(G,q,s,w)$ for several families of
graphs.

\section{Path Graph $L_n$}

The path graph $L_n$ is the graph consisting of $n$ vertices with each vertex
connected to the next one by one edge.  One may picture this graph as forming a
line, and in the physics literature this is commonly called a line graph.  We
use the alternate term ``path graph'' here because in mathematical graph theory
the line graph $L(G)$ of a graph $G$ refers to a different object (namely the
graph obtained by an ismorphism in which one maps the edges of $G$ to the
vertices of $L(G)$ and connects these resultant vertices by edges if the edges
of $G$ are connected to the same vertex of $G$).  For $n \ge 2$, the chromatic
number is $\chi(L_n)=2$. In \cite{phs} we gave some illustrative calculations
of $Z(L_n,q,s,v,w)$.  Here we present a general formula for this partition
function. Let
\beq
T_{Z,1,0}= \left( \begin{array}{cc}
    q-s+v & \ sw \\
    q-s   & \ w(s+v) \end{array} \right )
\label{TTsq10}
\eeq
\beq
H_{1,0}= \left( \begin{array}{cc}
    1 & 0 \\
    0 & sw \end{array} \right )
\label{H10}
\eeq
\beq
 \omega_1 = {q-s \choose 1}
\label{u1}
\eeq
and
\beq
s_1 = {1 \choose 1} \ .
\label{s1}
\eeq
Then
\beq
Z(L_n,q,s,v,w) = \omega_1^T \, H_{1,0} \, (T_{Z,1,0})^{n-1} \, s_1
\label{zline}
\eeq
and $Ph(L_n,q,s,w)=Z(L_n,q,s,-1,w)$.  It is straightforward to verify that 
our result for $Z(L_n,q,s,v,w)$ satisfies the relation (\ref{zscaled}). 
We note that 
\beq
{\rm det}(T_{Z,1,0}) = v(q+v)w \ , 
\label{dett}
\eeq
independent of $s$, and 
\beq
{\rm Tr}(T_{Z,1,0}) = q-s+v+w(s+v) \ . 
\label{tracet}
\eeq
The eigenvalues of $T_{Z,1,0}$ are the same as the eigenvalues with
coefficients of degree $d=0$ for the circuit graph $C_n$ given in Eqs. 
(5.3) of Ref. \cite{phs}, namely 
\beqs
\lambda_{Z,1,0,\pm} & = & \frac{1}{2} \Bigg [ q-s+v + w(s+v) \pm
 \Big [ \{q-s+v + w(s+v)\}^2-4vw(q+v)\Big ]^{1/2} \ \Bigg ] \ . 
\cr\cr
& & 
\label{lamcnplusminus}
\eeqs
Thus, we can also write 
\beq
Z(C_n,q,s,v,w) = {\rm Tr}[(T_{Z,1,0})^n] + (s-1)(vw)^n + (q-s-1)v^n \ .
\label{zcn}
\eeq

The graphs $L_n$, $C_n$, and, more generally, lattice strip graphs of some
transverse width $L_y$ and length $L_x=m$ are examples of recursive families of
graphs, i.e., graphs $G_m$ that have the property that $G_{m+1}$ can be
constructed by starting with $G_m$ and adding a given graph $H$ or, if
necessary, cutting and gluing in $H$.  For these graphs, $Z(G_m,q,s,v,w)$ has
the structure of a sum of coefficients that are independent of the length $m$
multiplied by $m$'th powers of some algebraic functions.  The results for
transfer matrices for the case $s=1$ in Ref. \cite{zth} elucidated this
structure for $s=1$, and our calculation of $Z(C_n,q,s,v,w)$ in Ref. \cite{phs}
and $Z(L_n,q,s,v,w)$ here elucidate this structure for general $s$.  Note that,
by Eq. (\ref{zterms}), the term in $Z(C_n,q,s,v,w)$ of highest order in $v$ is
$v^nu_n=[q+s(w-1)\sum_{j=0}^{n-1}w^j \ ] v^n$, part of which gives rise to the
last two terms in Eq. (\ref{zcn}).  We note that for $s=0$ or $w=1$, one can
check that our expressions for $Z(L_n,q,s,v,w)$ and $Z(C_n,q,s,v,w)$ simplify,
respectively, to $Z(L_n,q,v)=q(q+v)^{n-1}$ and
$Z(C_n,q,v)=(q+v)^n+(q-1)v^n$. Going from the case of $sw(w-1)=0$ to $sw(w-1)
\ne 0$, $Z(L_n,q,s,v,w)$ expands from a sum of one power to a sum involving two
powers, and $Z(C_n,q,s,v,w)$ expands from a sum of two powers to a sum of four
powers.

As our exact solutions for $Z(L_n,q,s,v,w)$ and $Z(C_n,q,s,v,w)$ show, the
field-dependent Potts partition functions $Z(G,q,s,v,w)$ do not, in general,
have any common factor. This contrasts with the case of the zero-field Potts
partition function, which always has an overall factor of $q$.  Similarly, in
the $v=-1$ special case defining the set-weighted chromatic polynomial, the
resultant polynomials $Ph(G,q,s,w)$ do not, in general, have a common factor.
For special values of $s$, $Ph(G,q,s,w)$ may reduce to a form with a common
factor.  The case $s=0$ (and the case $w=1$) for which this reduces to the
conventional chromatic polynomial is well-known; in this case $P(G,q)$ has, as
a common factor, $\prod_{j=0}^{\chi(G)-1}(q-j)$. Similarly, for $s=q$,
$Ph(G,q,q,w)$ has this common factor multipled by $w^n$. For the
special case $s=1$ and for a connected graph $G$ with at least one edge, it was
shown in Ref. \cite{ph} that $Ph(G,q,1,w)$ contains a factor $(q-1)$.  However,
it is not true that for a special case such as $s=2$, a connected graph $G$
with at least one edge contains a factor of $(q-s)$.  For example, using the
elementary result
\beq
Z(L_2,q,s,v,w) = s(s+v)w^2 + 2s(q-s)w +(q-s)(q-s+v) \ , 
\label{zl2}
\eeq
one sees that $Ph(L_2,q,1,w)=(q-1)(q-2+2w)$, but $Ph(L_2,q,2,w)=2w^2+
4(q-2)w+(q-2)(q-3)$, which has no common factor.

\section{Complete Graphs $K_n$}

The complete graph $K_n$ is the graph with $n$ vertices such that each vertex
is connected to every other vertex by one edge. The chromatic number is
$\chi(K_n)=n$ and the number of edges is $e(K_n)={n \choose 2}$. The
(conventional, unweighted) chromatic polynomial is 
\beq
P(K_n,q) = \prod_{j=0}^{n-1}(q-j) \ . 
\label{pkn}
\eeq
For our later calculations, we will need our previous result for 
$Ph(K_n,q,s,w)$ from Ref. \cite{phs}, which we mention here. 
We introduce a symbol $x_\theta \equiv x \theta(x)$, where
$\theta(x)$ is the step function from ${\mathbb R} \to \{0,1\}$ defined
as $\theta(x)=1$ if $x > 0$ and $\theta(x) = 0$ if $x \le 0$.  Our result is
\cite{phs} 
\beq
Ph(K_n,q,s,w) = \sum_{\ell=0}^n \beta_{K_n,\ell}(q,s) \, w^\ell
\label{phkn}
\eeq
where
\beq
\beta_{K_n,\ell}(q,s) = 
{n \choose \ell} \, \Big [ \prod_{i=0}^{(\ell-1)_\theta}
  (s-i) \Big ] \Big [ \prod_{j=0}^{(n-\ell-1)_\theta} (q-s-j) \Big ] \ .
\label{betaknell}
\eeq
Here it is understood that if the upper index on either of the two products in
Eq. (\ref{betaknell}) is negative, that product is absent, so that the 
first product is absent for $\ell=0$ and the second one is absent for $\ell=n$.
Note that 
\beq
\beta_{K_n,\ell}(q,s) = \beta_{K_n,n-\ell}(q,q-s) \ , 
\label{betakn_sym}
\eeq
in agreement with the general symmetry (\ref{phbetasym}).  Substituting this in
Eq. (\ref{phkn}) shows explicitly that our result for $Ph(K_n,q,s,w)$ satisfies
the symmetry relation (\ref{phsym}).  Note that $K_n$ is not a recursive family
of graphs, so one does not expect $Ph(K_n,q,s,w)$ to have the form of a sum of
coefficients multiplied by powers of certain algebraic functions, and it does
not, in contrast to $Ph(G_n,q,s,w)$ for recursive families $G_n$ such as $C_n$
or $L_n$.

The calculation of $Ph(K_n,q,s,w)$ for the cases $K_1$ and $K_2 = L_2$ are
elementary. For $K_3=C_3$ our general formula (\ref{phkn}) yields 
\beq
Ph(K_3,q,s,w) = P(K_3,s)w^3 + 3s(s-1)(q-s)w^2 + 3s(q-s)(q-s-1)w + P(K_3,q-s)
\label{phk3}
\eeq
while for $K_4$ we have 
\beqs
Ph(K_4,q,s,w) & = & P(K_4,s)w^4 + 4s(s-1)(s-2)(q-s)w^3 + 6s(s-1)(q-s)(q-s-1)w^2
\cr\cr
& + & 4s(q-s)(q-s-1)(q-s-2)w + P(K_4,q-s) \ . 
\label{phk4}
\eeqs

\section{$p$-Wheel Graphs $Wh^{(p)}=K_p + C_{n-p}$}

The $p$-wheel graph $Wh^{(p)}_n$ is defined as
\beq
Wh^{(p)}_n=K_p + C_{n-p} \ , 
\label{pwheel}
\eeq
i.e., the join of the complete graph $K_p$ with the circuit graph $C_{n-p}$.
(Given two graphs $G$ and $H$, the join, denoted $G+H$, is defined as the graph
obtained by joining each of the vertices of $G$ to each of the vertices of
$H$). (Here and below, no confusion should result from the use of the symbol
$H$ for a graph and $H$ for the external field; the meaning will be clear from
context.)  The family of $Wh^{(p)}_n$ graphs is a recursive family. For $p=1$,
$Wh^{(1)}_n$ is the wheel graph. The central vertex can be regarded as forming
the axle of the wheel, while the $n-1$ vertices of the $C_{n-1}$ and their
edges form the outer rim of the wheel. This is well-defined for $n \ge 3$, and
in this range the chromatic number is $\chi(Wh_n)=3$ if $n$ is odd and
$\chi(Wh_n)=4$ if $n$ is even.  Although $K_p$ is not defined for $p=0$, we may
formally define $Wh^{(0)}_n \equiv C_n$.  For the zero-field case, i.e., for
the usual, unweighted chromatic polynomial and for an arbitrary graph $G$,
\beq
P(K_p+G,q) = P(K_p,q)P(G,q-p) = q_{(p)}P(G,q-p) \ , 
\label{pkpg}
\eeq
where $q_{(m)}$ is the falling factorial, defined as
\beq
q_{(m)} = \prod_{j=0}^{m-1}(q-j) \ . 
\label{fallingfactorial}
\eeq
This result is a consequence of the fact that in assigning colors to the $p$
vertices of $K_p$, one must use $p$ different colors, and then, because of the
join condition, one must select from the other $q-p$ colors to color the
vertices of $G$.  In particular, for $Wh^{(p)}$, this gives
\beqs
P(Wh^{(p)}_n,q) & = & P(K_p,q)P(C_{n-p},q-p) \cr\cr
                & = &
q_{(p)} \Big [ (q-1-p)^{n-p} + (q-1-p)(-1)^{n-p} \Big ] \ . 
\label{ppwheel}
\eeqs
Note that, for arbitrary $p$, this chromatic polynomial consists of the
prefactor times the sum of the $(n-p)$'th powers of $N_{Wh^{(p)},\lambda}=2$
terms. For $p=1$, this number can be seen to be the $L_y=1$ special case of a
general formula in Eq. (3.2.15) of Ref. \cite{dg} for the join of $K_1$ with a
width-$L_y$ cyclic strip.

For the weighted-set chromatic polynomial, we generalize this coloring method
as follows.  Consider first $K_1 + G$.  There are two possible types of choices
for the color to be assigned to the vertex of $K_1$.  One type is to choose
this color to lie in the set $I_s$.  There are $s$ ways to make this choice,
and each gets a weighting factor of $w$.  For each choice, one then performs
the proper coloring of the vertices of $G$ with the remaining $q-1$ colors, of
which only $s-1$ can be used from the set $I_s$; this is determined by
$Ph(G,q-1,s-1,w)$.  The second type of coloring is to choose the color assigned
to the $K_1$ vertex to lie in the orthogonal set $I_s^\perp$.  There are
$(q-s)$ ways to make this choice, and since this is not the weighted set, there
is no weighting factor of $w$.  For each such choice, one then performs the
proper coloring of the vertices of $G$ with the remaining $q-1$ colors, of
which all $s$ colors in the set $I_s$ are available, but only $q-s-1$ colors in
the orthogonal set $I_s^\perp$ are available.  This yields the result
\beq
Ph(K_1 + G,q,s,w) = swPh(G,q-1,s-1,w)+(q-s)Ph(G,q-1,s,w) \ . 
\label{phk1g}
\eeq
To calculate $Ph(K_p + G,q,s,w)$ for a given graph $G$, one first carries out
the proper coloring of $K_1 + G$, using the result (\ref{phk1g}).  One then
joins the next vertex of $K_p$ to $K_1+G$ to get $K_2 + G$, using the relation
$K_1+(K_r+G) = K_{r+1} + G$ and iteratively applies Eq. (\ref{phk1g}). One
continues in this manner to carry out the proper coloring of the full join
$K_p+G$. This yields
\beq
Ph(K_p+G,q,s,w) = \sum_{\ell=0}^p \beta_{K_p,\ell}(q,s) \,
Ph(G,q-p,s-\ell,w)w^\ell \ . 
\label{phkpg}
\eeq

Utilizing this coloring method, we have calculated $Ph(Wh^{(p)}_n,q,s,w)$
for arbitrary $n$.  Let us define
\beq
a(p,q,s,w) = q-s-(p+1)+(s-1)w = q-(p+1)+s(w-1)-w
\label{a_pwheel}
\eeq
and
\beq
\lambda_{Wh^{(p)},\ell,\pm}(q,s,w) = 
\frac{1}{2}\bigg [a(p,q,s-\ell,w)\pm [a(p,q,s-\ell,w)^2+4w(q-p-1) \, ]^{1/2} 
\bigg ] \quad {\rm for} \ 0 \le \ell \le p \ . 
\label{lam_pwheel_ell}
\eeq
We note that for $\ell=0$, these $\lambda_{Wh^{(p)},\ell,\pm}(q,s,w)$ are equal
to the $v=-1$ special case of $\lambda_{Z,1,0,j}$ given in Eq. (5.3) of our
earlier Ref. \cite{phs} for the circuit graph with the replacement of $q$ by
$q-p$ (and with $j=1,2$ corresponding to $\pm$ here).  This is in accord with
the fact that the effect of the join of $K_p$ with $G$ is that the proper
$q$-coloring of $G$ can only use $q-p$ of the original $q$ colors.  We define
two additional terms that do not depend on $q$ or $s$,
\beq
\lambda_{Wh^{(p)},2p+3} = -w 
\label{lam_pwheel_w}
\eeq
and
\beq
\lambda_{Wh^{(p)},2p+4} = -1 \ . 
\label{lam_m_pwheel}
\eeq
The total number of $\lambda$'s for $Ph(Wh^{(p)}_n,q,s,w)$ is thus
\beq
N_{Ph(Wh^{(p)}),\lambda} = 2(p+2) \ . 
\label{nphwheel}
\eeq
Note that in contrast to the unweighted chromatic polynomial of $Wh^{(p)}_n$,
where the number of $\lambda$'s, $N_{P(Wh^{(p)}),\lambda}=2$, is independent of
$p$, here this number depends on $p$. In terms of these quantities, we find,
for the weighted-set chromatic polynomial for $Wh^{(p)}_n$, the result
\beqs
& & Ph(Wh^{(p)}_n,q,s,w) = \sum_{\ell=0}^p \beta_{K_p,\ell}(q,s) 
\Bigg [ [\lambda_{Wh^{(p)},\ell,+}(q,s,w)]^{n-p} +
         [\lambda_{Wh^{(p)},\ell,-}(q,s,w)]^{n-p} \, \Bigg ] w^\ell \cr\cr
& + &  
\Bigg [ \sum_{\ell=0}^p \beta_{K_p,\ell}(q,s) \, (s-\ell-1)w^\ell \, \Biggr ]
(-w)^{n-p} \cr\cr
& + & 
 \Bigg [ \sum_{\ell=0}^p \beta_{K_p,\ell}(q,s) \, (q-s-p+\ell-1)w^\ell \,
\Biggr ] (-1)^{n-p} \ . \cr\cr
& & 
\label{phpwheel}
\eeqs
This formula applies for integer $p \ge 1$ and also for $p=0$ if one sets
$\beta_{K_p,\ell}(q,s) \equiv \delta_{\ell,0}$ for $p=0$.  It can be checked
that for $p=0$, Eq. (\ref{phpwheel}) reduces to our result for $Ph(C_n,q,s,w)$
given as the special $v=-1$ case of Eqs. (5.3)-(5.5) in Ref.  \cite{phs}.  It
can also be verified that for $p=1$ and $s=1$, Eq. (\ref{phpwheel}) reduces to
the result given for this case in Eqs. (3.30)-(3.32) in Ref. \cite{ph}.
Furthermore, since the graph $Wh^{(1)}_4 = K_1 + K_3 = K_4$, it follows that
$Ph(Wh^{(1)}_4,q,s,w) = Ph(K_4,q,s,w)$. The symmetry (\ref{phsym}) is realized
as follows: the summation on the first line of Eq. (\ref{phpwheel}) goes into
itself, while the sum of the expressions on the two subsequent lines of Eq.
(\ref{phpwheel}) transforms into itself with the replacement of $w$ by $w^{-1}$
in these expressions and the prefactor $w^n$ appearing overall. One could also
study $Z(Wh^{(p)}_n,q,s,v,w)$, but we have focused here on
$Ph(Wh^{(p)},q,s,w)$, since its calculation can be performed by combinatoric
methods associated with the proper $q$-coloring condition. We give some
explicit examples of set-weighted chromatic polynomials $Ph(Wh^{(p)}_n,q,s,w)$
obtained from our general formula (\ref{phpwheel}) in the first appendix.

Following our notation in Ref. \cite{phs} and earlier works, the $n \to \infty$
limit of a family of $n$-vertex graphs $G_n$ is denoted $\{G\}$ and the
continuous accumulation set of the zeros of $Ph(G_n,q,s,w)$ in the complex
$q$ plane is denoted ${\cal B}_q$.  For recursive families of graphs, this
locus is determined as the solution of the equality in magnitude of two (or
more) $\lambda$'s of dominant magnitude, as a function of $q$ (with other
variables held fixed).  The other loci ${\cal B}_v$, etc. are defined in an
analogous manner.  These loci are typically comprised of curves and possible
line segments.  For studies of the $n \to \infty$ limit of chromatic
polynomials and their generalization to weighted-set chromatic polynomials, the
locus ${\cal B}_q$ is of primary interest.  Depending on the family of graphs,
the locus ${\cal B}_q$ may or may not cross the real $q$ axis.  If it does
cross the real $q$ axis, we denote the maximum (finite) point at which it
crosses this axis as $q_c$. Extending our previous result for the $p=0$ case of
$\{G\} = \{Wh^{(p)}\}$ in Eq. (7.17) of Ref. \cite{phs}, we find the 
following result for general $p$:
\beqs
& & q_c = 2 + p + \frac{s(1-w)}{1+w} \quad {\rm for} \ \{G\}=\{Wh^{(p)}\} \
{\rm and} \ 0 \le w \le 1 \quad {\rm and} \ 1 \le s \le p+2 \ . \cr\cr 
& &
\label{qc}
\eeqs
Regarding connections of this general formula to previously determined special
cases, (i) for $s=0$ or $w=1$, this reduces to the result $q_c=2+p$ for the $n
\to \infty$ limit of the chromatic polynomial $P(Wh^{(p)},q)$ given in Eq. (22)
of Ref. \cite{wc}; (ii) for $p=0$, this reduces to the result for the $n \to
\infty$ of $Ph(C_n,q,s,w)$ given in Eq.  (7.17) of Ref. \cite{phs}, and (iii)
for $s=1$, this reduces to the result for the $n \to \infty$ limit of
$Ph(Wh^{(1)},q,1,w)$ given in Eq. (10.1) of Ref. \cite{ph} (with the obvious
notation change $\{C\} \to \{Wh\}$). For the relevant interval $0 \le w \le 1$,
the value of $q_c$ in Eq. (\ref{qc}) is (a) greater than the value $q_c=2+p$
for the unweighted chromatic polynomial; (b) a monotonically increasing
function of $s$ for fixed $w$ in this DFSCP interval; and (c) a monotonically
decreasing function of $w$. These properties are consequences of the greater
suppression of color values in the set $I_s$ as $w$ decreases in the DFSCP
interval, finally restricting the vertex coloring to use colors from the
orthogonal set $I_s^\perp$ as $w$ reaches 0.  Thus, as $w$ decreases from 1 to
0, $q_c$ increases continuously from $2+p$ to $2+p+s$.  In contrast, the
left-hand part of the boundary locus ${\cal B}_q$ changes discontinuously; as
$w$ decreases by an arbitrarily small amount below 1, the point on the left
where ${\cal B}_q$ crosses the real $q$ axis jumps discontinuously from $q=p$
to $q=p+s$.  This behavior is in agreement with the fact that in the two limits
$w=1$ and $w=0$, ${\cal B}_q$ is comprised, respectively, of the unit circle
centered at $q=1+p$ and the unit circle centered at $q=1+s+p$.  The change in
the nature of the locus for $s > 2+p$ follows via the corresponding 
generalization of the analysis in Ref. \cite{phs} to $p \ge 0$.

\section{Effect of Multiple Edges in a Graph}

Consider a loopless graph $G=(V,E)$. Replace each edge with $\ell$ edges
joining the same pair of vertices and denote the resultant graph as $G_{\ell
e}$. Then the following is a theorem:
\beq
Z(G_{\ell e},q,s,v,w) = Z(G,q,s,v_\ell,w), \quad {\rm where} \ 
v_\ell = (v+1)^\ell-1 \ . 
\label{zgell}
\eeq
Clearly, if $v=0$, then $Z(G,q,s,v,w)=(q-s+sw)^n$, independent of the edge set
$E$ of $G$.  Hence, in this case, the operation of replacing each edge by
$\ell$ copies of the edge has no effect on the partition function.  This is
seen at an analytic level via the property that if $v=0$, then also $v_\ell=0$
for any (positive integer) $\ell$.  Further, for $v=-1$, where $Z(G,q,s,v,w)$
reduces to the weighted-set chromatic polynomial $Ph(G,q,s,w)$, the proper
$q$-coloring constraint is the same regardless of whether a given edge is
replicated or not, so again the replication does not affect this polynomial.
In Eq. (\ref{zgell}), this follows because if $v=-1$, then also $v_\ell=-1$
for any (positive integer) $\ell$.  Combining these results, we note that 
\beq
v_\ell-v \quad {\rm contains \ the \ factor} \quad v(v+1) \ . 
\label{velldif}
\eeq
Consequently, for any graph $G$ with at
least one edge (so that the operation of edge replication is not vacuous) and
for positive integer $\ell$, 
\beq
Z(G_{\ell e},q,s,v,w) - Z(G,q,s,v_\ell,w) \quad {\rm contains \ the \ factor}
 \quad v(v+1) \ . 
\label{zelldif}
\eeq

\section{Effects of Deletion and Contraction of Edges}  
\label{delconsection}

As noted above in Sect. \ref{properties}, in general, neither $Z(G,q,s,v,w)$
nor $Ph(G,q,s,w)$ satisfies the respective deletion-contraction relation.  It
is of interest to investigate how these polynomials deviate from the
deletion-contraction relation.  A natural measure of this deviation for a graph
$G$ is \cite{phs}

\beq
[\Delta Z(G,e,q,s,v,w)]_{DCR}=Z(G,q,s,v,w)
-\Big [ Z(G-e,q,s,v,w)+vZ(G/e,q,s,v,w)\Big ] \ . 
\label{zdelcondif}
\eeq
We also define $[\Delta Ph(G,e,q,s,w)]_{DCR} \equiv 
[\Delta Z(G,e,q,s,-1,w)]_{DCR}$.  We showed that \cite{phs} 
\beq
[\Delta Z(G,e,q,s,v,w)]_{DCR} \quad {\rm contains \ \ the \ \ factor} \ \ 
svw(w-1) \ ,
\label{deltafactors}
\eeq
and hence $[\Delta Ph(G,e,q,s,w)]_{DCR}$ contains a factor of $sw(w-1)$. 
A particularly elegant general formula can be obtained for this deviation in
the case of the family of star graphs, $S_n$, i.e., graphs consisting of one
central vertex with $n-1$ other vertices, each of which is only connected to 
this central vertex.  We find (for the nontrivial range $n \ge 2$) 
\beq
[\Delta Z(S_n,q,s,v,w)]_{DCR} = svw(w-1)[q+s(w-1)+wv]^{n-2} \ . 
\label{deltadcr_star}
\eeq

\section{Cycle Measure}

In view of the relation (\ref{zscaled}), one can define the following 
function, which serves as a measure of the presence of cycles in a graph $G$:
\beq
[\Delta Z(G,q,s,v,w)]_{cycles} = Z(G,q,s,v,w) - s^n Z(G,q',1,v',w)  \ . 
\label{deltaz_cycles}
\eeq
where $q'$ and $v'$ were defined in Eq. (\ref{qvprime}).  Clearly, this
difference vanishes if $s=1$, so, since it is a rational function, 
\beq
[\Delta Z(G,q,s,v,w)]_{cycles} \quad {\rm contains \ \ the \ \ factor} \quad
(s-1) 
\label{cyclediffactor}
\eeq
In Ref. \cite{phs} we derived the result 
\beq
[\Delta Z(C_n,q,s,v,w)]_{cycles} = \frac{(s-1)u_nv^n}{s} = 
\frac{(s-1)(q-s+sw^n)v^n}{s} \ . 
\label{deltazcncycles}
\eeq
This reflects the fact that $C_n$ contains one cycle. 

Here we present another example of this difference function.  Let us define
a path graph with $n$ vertices and each edge replaced by $\ell$ edges joining
the same adjacent vertices as $L_{n,\ell}$.  Note that $L_{2,2} = C_2$. For
$L_{3,2}$ we calculate
\beqs
& & [\Delta Z(L_{3,2},q,s,v,w)]_{cycles} = \frac{(s-1)v^2}{s} \bigg [ 
s(2s^2+v^2s+4vs+v^2)w^3 \cr\cr
& + & 2s^2(q-s)w^2 + 2s^2(q-s)w + (q-s)(-2s^2+v^2s+4vs+2sq+v^2) \bigg ] \ . 
\label{deltal3loop2cycles}
\eeqs

\section{Use of $Z(G,q,s,v,w)$ and $Ph(G,q,s,w)$ to Distinguish Between 
Tutte-Equivalent and Chromatically Equivalent Graphs}
\label{distinguish}

\subsection{General} 

Two graphs $G$ and $H$ are defined to be (i) chromatically equivalent if they
have the same chromatic polynomial, and (ii) Tutte-equivalent if they have the
same Tutte polynomial, or equivalently, zero-field Potts model partition
function.  Here the Tutte polynomial $T(G,x,y)$ of a graph $G$ is defined as
\beq
T(G,x,y) = \sum_{G' \subseteq G} (x-1)^{k(G')-k(G)}(y-1)^{c(G')} \ , 
\label{t}
\eeq
where $G'$ is a spanning subgraph of $G$ (and $c(G')$ and $k(G')$ were defined
above as, respectively, the number of linearly independent cycles and the
number of connected components of $G'$). This is equivalent to the
zero-field Potts model partition function, via the relation
\beq
Z(G,q,v) = (x-1)^{k(G)}(y-1)^n \, T(G,x,y) \ , 
\label{ztrel}
\eeq
where $y=v+1$ as in Eq. (\ref{kdef}) and $x=1+(q/v)$. The Tutte polynomial is
of considerable interest in mathematical graph theory, since it encodes much
information about a graph. However, although it distinguishes between many
graphs, there exist other pairs of graphs $G$ and $H$ that are different but
have the same Tutte polynomial.  An important property of our generalized
field-dependent Potts model partition function $Z(G,q,s,v,w)$ is that it can
distinguish between many Tutte-equivalent graphs.  Similarly, an important
property of the weighted-set chromatic polynomial is that it can distinguish
between many chromatically equivalent graphs.  We study this further in this
section.  This property is true for all $w$ and $s$ values except the special
values $w=1$, $w=0$, $s=0$, and $s=q$, for which $Z(G,q,s,v,w)$ is reducible to
a zero-field Potts partition function (as well as the trivial case $v=0$) and
similarly for $Ph(G,q,s,w)$.  reducible to a chromatic polynomial.  In
Ref. \cite{phs} we proved that for any two Tutte-equivalent graphs $G$ and $H$,
\beq
Z(G,q,s,v,w) - Z(H,q,s,v,w) \quad {\rm contains \ the \ factor} \quad 
s(q-s)vw(w-1) \ . 
\label{zghdiff}
\eeq
In the following, we will generally phrase our analysis in terms of how 
the field-dependent Potts partition function distinguishes between 
Tutte-equivalent graphs; the special cases of the various expressions for 
$v=-1$ then show how the weighted-set chromatic polynomial distinguishes
between different chromatically equivalent graphs. 

\subsection{Tree Graphs} 

A class of Tutte-equivalent (and, hence also chromatically equivalent) graphs
of particular interest is comprised of tree graphs, generically denoted $T_n$.
For these, $T(T_n,x,y)=x^{n-1}$, so
\beq
Z(T_n,q,v) = q(q+v)^{n-1} \quad {\rm and} \quad P(T_n,q) = q(q-1)^{n-1} \ .
\label{ztn}
\eeq
Note that $e(T_n)=n-1$ (and a tree graph cannot have any multiple edges).
There is only one tree graph with $n=1$ vertex, one with $n=2$ vertices, and
one with $n=3$ vertices.  There are two different tree graphs with $n=4$
vertices, namely the path graph, $L_4$, and the star graph, $S_4$.
Enumerations of tree graphs with larger numbers of vertices are given, e.g., in
Refs. \cite{harari,atlas}. Let us consider two different $n$-vertex tree graphs
(which thus have $n\ge 4$), denoted $G_t$ and $H_t$.  Since these have the same
number of edges, inspection of the general Eq. (\ref{zterms}) shows that for
the difference $Z(G_t,q,s,v,w)-Z(H_t,q,s,v,w)$, not only the $v^0$ and $v^n$
terms, but also the $v^1$ terms cancel.  Hence,
\beq
Z(G_t,q,s,v,w)-Z(H_t,q,s,v,w) \quad {\rm contains \ the \ factor} \ v^2  \ . 
\label{zghvsq}
\eeq

We recall that $I_s \subseteq I_q$, so that $0 \le s \le q$, and that $w \ge
0$, as follows for any physical field $H$.  These properties will be understood
implicitly in the following.  As preparation for the derivation of an
inequality concerning $Z(G_t,q,s,v,w)$ for $S_n$ and $L_n$ graphs, it is useful
to give some explicit examples.  Let us consider the two tree graphs with $n=4$
vertices, namely $S_4$ and $L_4$.  In the following, we will usually omit the
arguments $q,s,v,w$ in $Z(G,q,s,v,w)$ for brevity of notation. We have given
exact expressions for $Z(S_n)$ in Eq. (3.5) of Ref. \cite{phs} and for
$Z(L_n)$ in Eq. (\ref{zline}) above.  For our present purposes, we focus on the
expressions in terms of the spanning subgraph expansion. For $S_4$, this is
\beq
Z(S_4) = u_1^4 + 3vu_2u_1^2 + 3v^2 u_3u_1 + v^3u_4 \ , 
\label{zstar4}
\eeq
while for $L_4$ we have 
\beq
Z(L_4) = u_1^4 + 3vu_2u_1^2 + v^2(2u_3u_1+u_2^2) + v^3u_4 \ . 
\label{zline4}
\eeq
The difference in the structure of the term proportional to $v^2$ arises from
the differences in the spanning subgraphs with two edges in $S_4$ and
$L_4$.  Hence, 
\beq
Z(S_4)-Z(L_4) = v^2(u_3u_1 - u_2^2) = v^2s(q-s)w(w-1)^2 \ . 
\label{zstar4_minus_zline4}
\eeq
Since the last expression will appear as a factor in the differences 
$Z(G_t)-Z(H_t)$ to be presented below, we give it a symbol:
\beq
\mu \equiv s(q-s)v^2w(w-1)^2 
\label{mu} 
\eeq
and note that 
\beq
\mu \ge 0 \ , 
\label{mupositive}
\eeq
so that $Z(S_4)-Z(L_4) \ge 0$. 

There are three different tree graphs with $n=5$ vertices: $S_5$, $L_5$, and a
graph that we denote as $Y_5$, which has the form of a $Y$, with the vertical
part made up of three vertices and two edges (shown in Fig. 1 of Ref.
\cite{ph}).  The graph $Y_n$ is the generalization of this graph in which the
vertical part is comprised of $n-2$ vertices forming a path graph $P_{n-2}$ (so
that $Y_4 = S_4$). The spanning subgraph expansions for these graphs, in order
of decreasing maximal vertex degree, are
\beqs
& & Z(S_5) = u_1^5 + 4vu_2u_1^3 + 6v^2 u_3u_1^2 + 4v^3u_4u_1 + v^4u_5 \ , 
\cr\cr
& & 
\label{zstar5}
\eeqs
\beqs
& & Z(Y_5) = u_1^5 + 4vu_2u_1^3 + 2v^2(2u_3u_1^2+u_2^2u_1) 
+ v^3(3u_4u_1+u_3u_2) + v^4u_5 \ , 
\label{zy5}
\eeqs
and
\beqs
& & Z(L_5) = u_1^5 + 4vu_2u_1^3 + 3v^2( u_3u_1^2 + u_2^2u_1) 
+ 2v^3(u_4u_1 + u_3u_2) + v^4u_5 \ . 
\label{zline5}
\eeqs
Thus, for the differences, we have
\beqs
Z(S_5)-Z(Y_5) & = & 2v^2(u_3u_1^2 - u_2^2u_1) + 
                                     v^3(u_4u_1-u_3u_2) \cr\cr
              & = & \mu [2u_1+v(w+1)] \ , 
\label{zstar5_minus_zy5}                         
\eeqs
\beqs
Z(S_5)-Z(L_5) & = & 3v^2(u_3u_1^2-u_2^2u_1)+2v^3(u_4u_1-u_3u_2) \cr\cr
              & = & \mu [3u_1+2v(w+1)] \ , 
\label{zstar5_minus_zline5}
\eeqs
and
\beqs
Z(Y_5)-Z(L_5) & = & v^2(u_3u_1^2-u_2^2u_1) + v^3(u_4u_1-u_3u_2)
\cr\cr
                              & = & \mu[ u_1 + v(w+1)] \ . 
\label{zy5_minus_zline5}
\eeqs
Now (remembering that $0 \le s \le q$ and $w \ge 0$), for the ferromagnetic
range $v \ge 0$, for nonnegative $a$ and $b$, one has
\beq
au_1 + bv(w+1) \ge 0 \ . 
\label{abpos}
\eeq
Hence, for the ferromagnetic case, each of the differences $Z(S_5)-Z(Y_5)$,
$Z(S_5)-Z(L_5)$, and $Z(Y_5)-Z(L_5)$ is non-negative.

From these explicit examples, one sees that the origin of these inequalities
can be traced to inequalities among products of the $u_r$'s.  We proceed to
prove two lemmas and then a general theorem.  Our first lemma is 
\beq
u_{n-1} \, u_1 \ge u_{n-\ell} \, u_\ell \quad {\rm for} \ n \ge 2 \ {\rm and}
 \ 2 \le \ell \le n-2 \ . 
\label{urel1}
\eeq
To verify this lemma, we expand and factor the given expression: 
\beqs 
u_{n-1} \, u_1 - u_{n-\ell} \, u_\ell & = & 
          s(q-s)w(1+w^{n-2}-w^{\ell-1}-w^{n-\ell-1}) \cr\cr
    & = & s(q-s)w(w^{n-\ell-1}-1)(w^{\ell-1}-1) \cr\cr
    & = & s(q-s)w(w-1)^2\Big [ \sum_{i=0}^{n-\ell-2} w^i \Big ] 
                        \Big [ \sum_{j=0}^{\ell-2} w^j \Big ] \ge 0 \ . 
\label{urel1calc}
\eeqs
This lemma shows that the difference $u_3u_1 - u_2^2$ that appears multiplying
$v^2$ in Eqs. (\ref{zstar4_minus_zline4}, (\ref{zstar5_minus_zy5}), 
(\ref{zstar5_minus_zline5}), and (\ref{zy5_minus_zline5}) is nonnegative, and
similarly that the difference $u_4u_1-u_3u_2$ that appears multiplying $v^3$ in
the last three of these equations is nonnegative.  

Differences of the form $Z(G_t)-Z(H_t)$ for higher values of $n$ involve
differences of higher products of $u_r$ factors, and there is an analogous
inequality for these products.  We prove this as a second lemma.  Let us
consider a generic term in Eq. (\ref{clusterws}), for the spanning subgraph $G'
= \oplus G'_i$ with $k(G')$ connected components, $G'_i$, each with $n(G'_i)$
vertices.  This has the form (\ref{uform}) satisfying the relation
(\ref{sum_ngprime_i}).  Our second lemma is, with $\ell=n-k(G')+1$, 
\beq
u_\ell \, u_1^{n-\ell} \ge \prod_{j=1}^{k(G')} \, u_{n(G'_j)} \quad {\rm for}
\quad n \ge 2 \quad {\rm and} \quad 1 \le \ell \le n, \ \ i.e., \ 
1 \le k(G') \le n \ . 
\label{uugen}
\eeq
For example, for the case $n=6$, this lemma yields the inequalities $u_4 u_1^2
\ge u_3^2$, $u_4 u_1^2 \ge u_2^3$, and $u_4 u_1^2 \ge u_4 u_2$. This lemma is
proved by the same method as Lemma 1.

Combining the expression for $Z(S_n,q,s,v,w)$ in the first line of
Eq. (\ref{zstar}) with our other results above, we have the following 
theorem: For the ferromagnetic case, 
\beq
Z(S_n,q,s,v,w) - Z(T_n,q,s,v,w) \ge 0 \quad {\rm for} \ v \ge 0
\label{zstar_minus_ztree_pos}
\eeq
for any tree graph $T_n$. This is proved by applying the two lemmas above to
the terms in the spanning subgraph expansions of these partition functions for
$S_n$ and a generic tree graph $T_n$.  In the second appendix we give further
explicit results for differences of field-dependent partition functions for
tree graphs with $n=6$ vertices.

The difference $Z(Y_5)-Z(L_5)$ in
Eq. (\ref{zy5_minus_zline5}) (where we omit the arguments for brevity of
notation) can also be understood using the recursive
relation for $n \ge 5$: 
\beqs
& & Z(Y_n)-Z(L_n) = \sum_{j=1}^{n-4}v^{j-1}u_j [ Z(Y_{n-j})-Z(L_{n-j}) ] 
\cr\cr
& & + v^{n-4}(\sum_{j=0}^{n-4}w^j) [ Z(Y_4)-Z(L_4) ] \ , 
\label{zyn_minus_zlinen}
\eeqs
where $Z(Y_4)-Z(L_4)=Z(S_4)-Z(L_4)=\mu$ was given in
Eq. (\ref{zstar4_minus_zline4}).  For the ferromagnetic range $v \ge 0$, each
term on the right-hand side of Eq. (\ref{zyn_minus_zlinen}) is nonnegative, and
hence this proves the inequality 
\beq
Z(Y_n,q,s,v,w)-Z(L_n,q,s,v,w) \ge 0 \quad {\rm for} \ \ v \ge 0 \ . 
\label{zyn_minus_zlinen_pos}
\eeq
Combining (\ref{zstar_minus_ztree_pos}) and (\ref{zyn_minus_zlinen_pos}), we
have
\beq
Z(S_n,q,s,v,w) \ge Z(Y_n,q,s,v,w) \ge Z(L_n,q,s,v,w) \quad {\rm for} 
\ \ v \ge 0 \ . 
\label{zstar_minus_zyn_zline}
\eeq

\subsection{Properties of Graphs Intersecting in a Complete Graph} 

One class of chromatically equivalent graphs consists of graphs whose chromatic
polynomials can be shown to be equal by an application of the complete graph
intersection theorem.  We recall this theorem.  Let us consider a graph $G$
that has the property of being composed of the union of two subgraphs, $G = G_1
\cup G_2$, such that $G_1 \cap G_2 = K_m$ for some $m$.  In the rest of this
subsection, we assume that $G$ has this property.  Then $P(G,q)$ satisfies the
relation
\beq
P(G,q) = \frac{P(G_1,q)P(G_2,q)}{P(K_m,q)} \ .
\label{intersectiontheorem}
\eeq
This is sometimes called the complete-graph intersection theorem (KIT) for
chromatic polynomials.  In contrast, in general, $Ph(G,q,s,w)$ is not equal to
$Ph(G_1,q,s,w)Ph(G_2,q,s,w)/Ph(K_m,q,s,w)$.  This equality holds only for the
four values $w=1$, $w=0$, $s=0$, and $s=q$ where $Ph(G,q,s,w)$ reduces to a
chromatic polynomial.  As a measure of the deviation from equality, we define
\beq
[\Delta Ph(G,q,s,w)]_{KIT} \equiv Ph(G,q,s,w) -
\frac{Ph(G_1,q,s,w)Ph(G_2,q,s,w)}{Ph(K_m,q,s,w)} \ .
\label{kit_diff}
\eeq

Let us consider a graph with $n=5$ vertices, denoted $G_{LKL}$, comprised of a
triangle $K \equiv K_3$ with two line segments $L$, each of length one edge,
emanating outward from two vertices of the triangle. This graph LKL has $n=5$,
$e=5$ and $c=1$.  A second graph, $G_{KLL}$, also with $n=5$, $e=5$ and $c=1$,
is comprised of a triangle $K_3$ with a line segment two edges long emanating
outward from one vertex of the triangle.  These graphs are Tutte-equivalent,
with $\triangle$ 
\beq
T(G_{LKL},x,y) = T(G_{KLL},x,y) = x^2(x+x^2+y) \ , 
\label{tuttepoly_lkl}
\eeq
or equivalently, 
\beq
Z(G_{LKL},q,v) = Z(G_{KLL},q,v) = q(q+v)(q^2+3qv+3v^2+v^3) \ . 
\label{z0lkl}
\eeq
It follows that these graphs are also chromatically equivalent, with chromatic
polynomial
\beq
P(G_{LKL},q) = P(G_{KLL},q) = q(q-1)^3(q-2) \ . 
\label{chrompoly_lkl}
\eeq
In contrast, the field-dependent Potts partition function and the weighted-set
chromatic polynomial successfully distinguish between these graphs.  For
the LKL graph we calculate
\beqs
& & Z(G_{LKL},q,s,v,w)=Z(G_{LKL},s,v)w^5+s(q-s)(s+v)(5s^2+10sv+7v^2+2v^3)w^4 
\cr\cr
& + & s(q-s)\bigg [ (q-s+2)v^3+(7q-s)v^2+10s^2(q-s-v)+15sqv \bigg ]w^3 \cr\cr
& + & s(q-s)\bigg [ (s+2)v^3+(6q+s)v^2+10(q-s)^2(s-v)+15(q-s)qv \bigg ]w^2
\cr\cr
& + & s(q-s)(q-s+v)\bigg [ 5(q-s)^2+10(q-s)v+7v^2+2v^3 \bigg ] w \cr\cr
               & + & Z(G_{LKL},q-s,v) 
\label{zglkl}
\eeqs
where $Z(G_{LKL},q,v)=Z(G_{KLL},q,v)$ was given above in Eq. (\ref{z0lkl}). 
For the $G_{KLL}$ graph we calculate 
\beqs
& & Z(G_{KLL},q,s,v,w) = Z(G_{KLL},s,v)w^5 \cr\cr
& + & s(q-s)\bigg [ 5s^3+15s^2v+16sv^2+2sv^3+5v^3+v^4 \bigg ] w^4 \cr\cr
& + & s(q-s)\bigg [ -10s^3+10s^2(q-v)+sv(15q+2v-v^2)+6qv^2+qv^3+4v^3+v^4
\bigg ]w^3 \cr\cr
& + & s(q-s)\bigg [ -10(q-s)^3+10(q-s)^2(q-v)+(q-s)v(15q+2v-v^2) \cr\cr
& + & 6qv^2+qv^3+4v^3+v^4 \bigg ]w^2 \cr\cr
& + & s(q-s)\bigg [ 5(q-s)^3+15(q-s)^2v+16(q-s)v^2+2(q-s)v^3+5v^3+v^4 \bigg ]w
 \cr\cr
& + & Z(G_{KLL},q-s,v) \ . 
\label{zgkll}
\eeqs
Thus, $Z(G_{LKL},q,s,v,w)$ is not, in general, equal to $Z(G_{KLL},q,s,v,w)$
and, taking the $v=-1$ special case, $Ph(G_{LKL},q,s,w)$ is 
not, in general, equal to $Ph(G_{KLL},q,s,w)$.  For the differences, we find 
\beqs
& & Z(G_{LKL},q,s,v,w)-Z(G_{KLL},q,s,v,w) = \cr\cr
& &  \mu \bigg [ s(w-1)+(1+w)v^2+2wv+q+2v \bigg ]
\label{zglkl_minus_zgkll}
\eeqs
and thus 
\beqs
& & Ph(G_{LKL},q,s,w)-Ph(G_{KLL},q,s,w) = s(q-s)w(w-1)^2
\bigg [ s(w-1)+q-1-w \bigg ] \ . \cr\cr
& & 
\label{delta_phglkl_phgkll}
\eeqs
These calculations provide another illustration of how $Z(G,q,s,v,w)$ can
distinguish between different graphs that yield the same Tutte polynomial, and
how $Ph(G,q,s,w)$ can distinguish between different graphs that yield the same
chromatic polynomial.

In the context of graphs that can be decomposed into the union of subgraphs
that intersect in a complete graph, it is also useful to calculate $Z$ and $Ph$
for the graph consisting of two $K_3$'s meeting at a common vertex, $\bowtie$,
denoted $G_{KK}$.  This graph has $n=5$, $e=6$, and $c=2$.  The Tutte
polynomial is $T(G_{KK},x,y) = (x+x^2+y)^2$, or equivalently
\beq
Z(G_{KK},q,v) = \frac{Z(K_3,q,v)^2}{Z(K_1,q,v)} = q(q^2+3qv+3v^2+v^3)^2
\label{zgkk}
\eeq
so that
\beq
P(G_{KK},q) = \frac{P(K_3,q)^2}{P(K_1,q)} = q(q-1)^2(q-2)^2
\label{pgkk}
\eeq
Note that these polynomials factorize.  This is not the case with 
$Z(G_{KK},q,s,v,w)$ and $Ph(G_{KK},q,s,w)$.  We calculate
\beqs
& & Z(G_{KK},q,s,v,w) = Z(G_{KK},s,v)w^5 + 
s(q-s)(s+v)(5s^2+13sv+12v^2+4v^3)w^4 \cr\cr
& + & 2s(q-s)\bigg [-5s^3+s^2(5q-6v)+sv(9q-v^2)+5qv^2+(q+3)v^3+v^4 \bigg ]w^3 
\cr\cr
& + & 2s(q-s)\bigg [ -5(q-s)^3+(q-s)^2(5q-6v)+(q-s)v(9q-v^2)
+ 5qv^2 + (q+3)v^3+v^4 \bigg ]w^2 \cr\cr
& + & s(q-s)(q-s+v)\bigg [5(q-s)^2+13(q-s)v+12v^2+4v^3 \bigg ]w \cr\cr
& + & Z(G_{KK},q-s,v) 
\label{zgkk_sw}
\eeqs
and hence
\beqs
& & Ph(G_{KK},q,s,w) = P(G_{KK},s)w^5 + s(q-s)(s-1)^2(5s-8)w^4 \cr\cr
& + & 2s(q-s)(s-1)\bigg [ 5s(q-s)+s-4q+2 \bigg ]w^3 + 
    2s(q-s)(q-s-1)\bigg [ 5s(q-s)+(q-s)-4q+2 \bigg ]w^2 \cr\cr
& + & s(q-s)(q-s-1)^2\bigg [ 5(q-s)-8 \bigg ]w + P(G_{KK},q-s) 
\label{phgkk}
\eeqs
This example thus further illustrates how the factorization properties of the
field-dependent Potts model partition function and set-weighted chromatic
polynomial differ from those of the zero-field Potts model partition function
and (unweighted) chromatic polynomial.

\section{Conclusions}

In this paper we have presented exact results on $Z(G,q,s,v,w)$, the partition
function of the Potts model in an external generalized magnetic field that
favors or disfavors spin values in a subset $I_s$ of the full set $I_q$ on
various families of graphs $G$, and on $Ph(G,q,s,w)$, the weighted-set
chromatic polynomial. In particular, we have presented new general calculations
of $Z(G,q,s,v,w)$ for the case of path (line) graphs $L_n$ and
$Ph(Wh^{(p)}_n,q,s,w)$ for $p$-wheel graphs $Wh^{(p)}_n$.  We have discussed
various features of our exact results for path, line, circuit, star, and
complete graphs.  We have derived powerful new upper and lower bounds on
$Z(G,q,s,v,w)$ in terms of zero-field Potts partition functions with certain
transformed arguments. We have also proved inequalities for the field-dependent
Potts partition function on different families of tree graphs.  An important
property of $Z(G,q,s,v,w)$ is the fact that it can distinguish between
Tutte-equivalent graphs, and similarly, $Ph(G,q,s,w)$ can distinguish between
certain chromatically equivalent graphs.  We have elucidated this property with
our inequalities and with explicit calculations.  Some new results for
quantities such as $f(\{G\},q,s,v,w)$, $\Phi(\{G\},q,s,w)$, and $q_c$ defined
in the $n \to \infty$ limits of recursive graph families have also been given.

There are a number of interesting directions for future study.  One direction
is to investigate additional graph-theoretic applications of $Z(G,q,s,v,w)$ and
$Ph(G,q,s,w)$.  Second, it is clearly valuable to calculate $Z(G,q,s,v,w)$ and
$Ph(G,q,s,w)$ for other families of graphs, in particular, recursive ones such
as lattice strips and study their properties.  We have described some of the
interesting differences in thermodynamic behavior between our model with a
generalized external magnetic field and the Potts model with a conventional
magnetic field. It would be worthwhile to explore these differences further for
various values of $q$, $s$, $H$, and temperature on lattices in $d \ge 2$
dimensions, using methods such as series expansions and Monte Carlo
simulations. 

\begin{acknowledgments}

R.S. thanks Prof. S.-C. Chang for valuable collaborations on
Refs. \cite{hl}-\cite{ph}. The present research was partly supported by the
grant NSF-PHY-06-53342.

\end{acknowledgments}

\newpage

\section{Appendix 1} 

In this appendix we present some explicit expressions for the set-weighted
chromatic polynomials $Ph(Wh^{(p)}_n,q,s,w)$ obtained from our general formula
(\ref{phpwheel}). We begin with case $p=1$, i.e., the family of wheel graphs,
$Wh^{(1)}_n$. The first nondegenerate case is $n=4$, for which $Wh^{(1)}_4 =
K_4$, as noted (cf. Eq. (\ref{phk4})).  For $n=5$, we have
\beqs
& & Ph(Wh^{(1)}_5,q,s,w) = s(s-1)(s-2)(s^2-5s+7)w^5 + 
s(s-1)[5s^2-19(s-1)](q-s)w^4 \cr\cr
& + & 2s(s-1)(q-s)(5qs-7q-5s^2+3s+6)w^3 \cr\cr
& + & 2s(q-s)(q-s-1)(-4q+5qs-5s^2+6-3s)w^2 \cr\cr
& + & s(q-s)(q-s-1)[5(q-s)^2-19(q-s-1)]w \cr\cr
& + & (q-s)(q-s-1)(q-s-2)[(q-s)^2-5(q-s)+7] \ . 
\label{ph1wheeln5}
\eeqs
For $n=6$, 
\beqs
& & 
Ph(Wh^{(1)}_6,q,s,w) = s(s-1)(s-2)(s-3)(s^2-4s+5)w^6 \cr\cr
& + & 2s(s-1)(s-2)[3s^2-11(s-1)](q-s)w^5 \cr\cr
& + & 5s(s-1)(q-s)(3qs^2-9qs+7q-3s^3+7s^2-s-5)w^4 + 
20s(s-1)^2(q-s)(q-s-1)^2w^3 \cr\cr
& + & 5s(q-s)(q-s-1)(3q^2s-2q^2+6q-6qs^2-5qs+3s^3+7s^2-5+s)w^2 \cr\cr
& + & 
2s(q-s)(q-s-1)(q-s-2)[(3(q-s)^2-11(q-s-1)]w \cr\cr
& + & (q-s)(q-s-1)(q-s-2)(q-s-3)[(q-s)^2-4(q-s)+5] \ . 
\label{ph1wheeln6}
\eeqs

For the family of $p$-wheel graphs with $p=2$, i.e., $Wh^{(2)}_n$, we first
note that $Wh^{(2)}_4 = K_2 + K_2 = K_4$ and $Wh^{(2)}_5 = K_2 + K_3 = K_5$,
both of which are subsumed by our general formula (\ref{phkn}). For
$Wh^{(2)}_6$ our general result (\ref{phpwheel}) yields
\beqs
& & Ph(Wh^{(2)}_6,q,s,w) = s(s-1)(s-2)(s-3)(s^2-7s+13)w^6 \cr\cr
& + & 2s(s-1)(s-2)(3s^2-17s+25)(q-s)w^5 \cr\cr
& + & s(s-1)(q-s)(15qs^2-63qs+67q-15s^3+50s^2-12s-59)w^4 \cr\cr
& + & 4s(s-1)(q-s)(q-s-1)[5s(q-s)-8q+13]w^3 \cr\cr
& + & s(q-s)(q-s-1)(15q^2s-13q^2-37qs+55q-30qs^2+12s+15s^3+50s^2-59)w^2 \cr\cr
& + & 2s(q-s)(q-s-1)(q-s-2)[3(q-s)^2-17(q-s)+25]w \cr\cr
& + & (q-s)(q-s-1)(q-s-2)(q-s-3)[(q-s)^2-7(q-s)+13] \ . 
\label{ph2wheeln6}
\eeqs
Explicit examples for higher values of $p$ and $n$ can be calculated in a
similar manner from our general formula (\ref{phpwheel}). 

\section{Appendix 2} 

\begin{figure}
\epsfxsize=6in
\epsffile{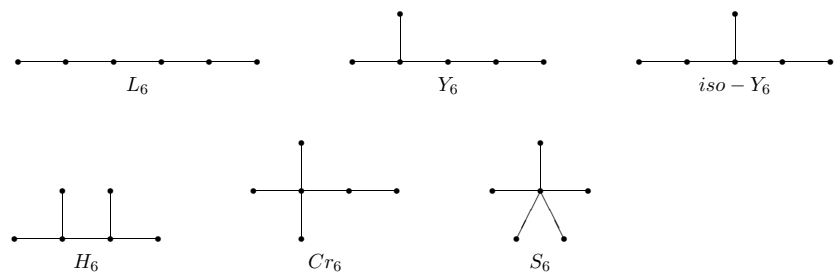}
\caption{Tree graphs with $n=6$ vertices.}
\label{tree6}
\end{figure}

As discussed in the text, an important property of $Z(G,q,s,v,w)$ is that it
can distinguish between different Tutte-equivalent graphs, and a similarly
important property of $Ph(G,q,s,w)$ is that it can distinguish between many 
chromatically equivalent graphs.  Tree graphs provide a basic context in which
to explore this property, since any two tree graphs are both chromatically
equivalent and Tutte-equivalent.  In the present text and in earlier work we
have given $Z(G,q,s,v,w)$ for tree graphs up to and including $n=5$
vertices. Here we give results for tree graphs with $n=6$ vertices.  There are
six such tree graphs, as shown in Fig. 6 of Ref. \cite{ph}, reproduced here as
Fig. \ref{tree6} for the reader's convenience, labelled (i) $L_n$, (ii) $Y_6$,
(iii) iso$Y_6$, (iv) $H_6$, (v) $Cr_6$, and (vi) $S_6$. These are listed in
order of increasing maximal vertex degree $\Delta$.  There are thus 
${6 \choose 2}=15$ differences of $Z(G,q,s,v,w)$ polynomials for these six
graphs.  We list these below, as differences with respect to the graph with
highest maximal vertex degree, $S_6$ first, then with respect to the graph with
next-highest maximal vertex degree, $Cr_6$, and so forth.  Furthermore, 
since the arguments are the same for all of the partition functions, we omit
them, writing $Z(G,q,s,v,w)-Z(H,q,s,v,w) \equiv Z(G)-Z(H)$.  The 
differences are 
\beqs
& & Z(S_6)-Z(Cr_6) = \mu \bigg [ (3s^2+3sv+v^2)w^2 \cr\cr
& + & \{6s(q-s)+3qv+v^2\}w + \{3(q-s)^2+3(q-s)v+v^2\} \bigg ]
\label{zstar6_minus_zcross6}
\eeqs
\beq
Z(S_6)-Z(H_6) = \mu [(2s+v)w + \{2(q-s)+v\}]^2 
\label{zstar6_minus_zyy6}
\eeq
\beqs
& & Z(S_6)-Z(IsoY_6) = \mu \bigg [ (5s^2+6sv+2v^2)w^2 \cr\cr
& + & \{10s(q-s)+5qv+2v^2\}w + \{5(q-s)^2+6(q-s)v+2v^2\} \bigg ] 
\label{zstar6_minus_zisoy6}
\eeqs
\beqs
& & Z(S_6)-Z(Y_6) = \mu \bigg [ (5s^2+6sv+2v^2)w^2 \cr\cr
& + & \{10s(q-s)+6qv+3v^2\}w + \{5(q-s)^2+6(q-s)v+2v^2\} \bigg ] 
\label{zstar6_minus_zy6}
\eeqs
\beqs
& & Z(S_6)-Z(L_6) = \mu \bigg [ (6s^2+8sv+3v^2)w^2 \cr\cr
& + & \{ 12s(q-s)+7qv+4v^2 \}w + \{ 6(q-s)^2+ 8(q-s)v+3v^2 \} \bigg ]
\label{zstar6_minus_zline6}
\eeqs
\beqs
& & Z(Cr_6)-Z(H_6) = \mu \Big [ q+s(w-1)+v \Big ] \Big [q+s(w-1)+vw \Big ] 
\cr\cr
& = & \mu \bigg [ s(s+v)w^2 + \{2s(q-s)+v(q+v)\}w + (q-s)(q-s+v) \bigg ]
\label{zcross6_minus_zyy6}
\eeqs
\beqs
& & Z(Cr_6)-Z(IsoY_6) = \mu \bigg [ (s+v)(2s+v)w^2 + (2s+v)\{2(q-s)+v\}w
\cr\cr & + & (q-s+v)\{2(q-s)+v\} \bigg ] 
\label{zcross6_minus_zisoy6}
\eeqs
\beqs
& & Z(Cr_6)-Z(Y_6) = \mu \Big [ 2\{q+s(w-1) \} + v(w+1) \Big ]
\Big [ q+s(w-1)+v(w+1) \Big ] \cr\cr
& = & \mu \bigg [ (s+v)(2s+v)w^2+ \{4s(q-s)+3qv+2v^2\}w +
(q-s+v)\{2(q-s)+v\} \bigg ] \cr\cr
& &
\label{zcross6_minus_y6}
\eeqs
\beqs
& & Z(Cr_6)-Z(L_6) = \mu \bigg [ (s+v)(3s+2v)w^2 + \{6s(q-s)+4qv+3v^2\}w 
\cr\cr
& + & (q-s+v)\{3(q-s)+2v\} \bigg ]
\label{zcross6_minus_zline6}
\eeqs
\beq
Z(H_6)-Z(IsoY_6) = \mu \bigg [ (s+v)^2w^2+ \{2s(q-s)+qv\}w+ (q-s+v)^2 \bigg ] 
\label{zyy6_minus_zisoy6}
\eeq
\beq
Z(H_6)-Z(Y_6) = \mu \bigg [ (s+v)^2w^2+ \{2s(q-s)+2qv+v^2 \}w + 
(q-s+v)^2 \bigg ]
\label{zyy6_minus_zy6}
\eeq
\beq
Z(H_6)-Z(L_6) = \mu \bigg [ 2(s+v)^2w^2+ \{4s(q-s)+3qv+2v^2 \}w + 
2(q-s+v)^2 \bigg ]
\label{zyy6_minus_zline6}
\eeq
\beq
Z(IsoY_6)-Z(Y_6)=\mu wv(q+v) = s(q-s)v^3(q+v)[w(w-1)]^2 
\label{zisoy6_minus_zy6}
\eeq
\beq
Z(IsoY_6)-Z(L_6)=\mu \Big [ q+s(w-1)+v(w+1) \Big ]^2 = 
\mu \Big [ (s+v)w + q-s+v \Big ]^2
\label{zisoy6_minus_zline6}
\eeq
\beq
Z(Y_6)-Z(L_6)=\mu \bigg [ (s+v)^2w^2 + \{2s(q-s)+v(q+v) \}w + (q-s+v)^2 \bigg ]
\label{zy6_minus_zline6}
\eeq
In addition to our theorems (\ref{zstar_minus_ztree_pos}) and
(\ref{zyn_minus_zlinen_pos}), we observe that (for $0 \le s \le q$ and $w
\ge 0$) all of these differences are non-negative for the ferromagnetic range
$v \ge 0$.

\newpage

\vfill
\eject
\end{document}